%% file: sample-journal.tex
 \documentclass[acmlarge, dvipsnames ]{acmart}

\usepackage{graphicx}
\usepackage{colortbl}
\usepackage{float} 
\usepackage{url} 
\usepackage{color}

\usepackage{subfig}

\usepackage{caption}
\usepackage{setspace}
\usepackage{adjustbox}  

\usepackage{graphicx}
\usepackage{float} 
\usepackage{url} 
\usepackage{color}

  \usepackage{graphicx}
    \usepackage{amssymb}
    \usepackage{latexsym}
    \usepackage{amsfonts}
    \usepackage{amsmath}
    \usepackage{graphics}
    \usepackage{setspace}
    \usepackage{graphicx}
    \usepackage{epstopdf}
    \usepackage{color}
    \usepackage{float}
    \usepackage{wrapfig}
    \usepackage{color}
    \usepackage{rotating}
    \usepackage{array} 
    \usepackage{dblfloatfix}
    \usepackage{breakurl}

    \usepackage{graphics}
    \usepackage{setspace}
    
    \usepackage{graphicx}
    \usepackage{epstopdf}  
    \usepackage{color}
    \usepackage{float}
    \usepackage{wrapfig}
    \usepackage{color}
    
    \usepackage{graphicx}
    \usepackage{caption}
    \usepackage{setspace}
    
      \usepackage{seqsplit}
  \usepackage{lipsum}
   \usepackage{hyperref}
   \hypersetup{colorlinks=false}

    \usepackage{algorithm}
    \usepackage{algorithmic}
    \usepackage{morefloats}
     \usepackage{textcomp}

       \usepackage{caption}
       \usepackage{setspace}
       \usepackage{adjustbox}  
    \usepackage{multirow}    
     \definecolor{Computation}{RGB}{255,255,255}
     \definecolor{Consent}{RGB}{255,255,255}
     \definecolor{Compliance}{RGB}{255,255,255}

     \definecolor{minimise}{RGB}{255,222,222}
     \definecolor{hide}{RGB}{253,228,205}
     \definecolor{seperate}{RGB}{250,250,202}
     \definecolor{aggregate}{RGB}{219,238,254}
     \definecolor{inform}{RGB}{237,215,253}
     \definecolor{controlenforce}{RGB}{251,214,253}     
     \definecolor{demostrate}{RGB}{228,255,227}

  \usepackage{colortbl}

\acmVolume{0}
\acmNumber{0}
\acmArticle{0}
\acmYear{0000}
\acmMonth{0}

\acmBadgeR[http://ctuning.org/ae/ppopp2016.html]{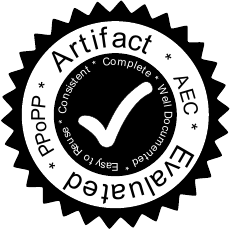}

\setcopyright{acmcopyright}

\acmDOI{0000001.0000001}

\begin{document}
\title{PizzaBox: Studying Internet Connected Physical Object Manipulation based Food Ordering\vspace{20pt}} 

\author{Luke Jones}
\affiliation{%
	\institution{Cardiff Univerisity}
	\country{UK}}

\author{Charith Perera}
\affiliation{%
	\institution{Cardiff Univerisity}
	\country{UK}}
%


\begin{abstract}
This paper presents the designing and testing of PizzaBox, a 3D printed, interactive food ordering system that aims to differ from conventional food ordering systems and provide an entertaining and unique experience when ordering a pizza by incorporating underlying technologies that support ubiquitous computing. The PizzaBox has gone through both low and medium fidelity testing while working collaboratively with participants to co-design and refine a product that is approachable to all age groups while maintaining a simple process for ordering food from start to finish. Final testing was conducted at an independent pizzeria where interviews with participants lead us to develop four discussion themes 1) usability and end user engagement, 2) towards connected real-time products and services, 3) healthy eating, 4) evolution of food ordering systems. Our interviews show that in general, PizzaBox would have a greater appeal to a younger audience by providing a fantasy of helping in the creation and baking of the pizza but also has a novelty value that all ages would enjoy. We investigate the effect that the PizzaBox has in encouraging new healthy habits or promoting a healthier lifestyle as well as how we can improve PizzaBox to better encourage these lifestyle changes. 


\end{abstract}

%
%
%

\ccsdesc[500]{Human-centered computing~Empirical studies in ubiquitous and mobile computing}
%
%


\keywords{Ubiquitous Computing, Internet of Things, Physical Objects Interaction, Connected Things}


\maketitle

\input{samplebody-journals}

\end{document}

%% file: samplebody-journals.tex

\section{Introduction}
\label{sec:Introduction}
We have all ordered pizza using either a web browser or a mobile application but can we do something different? Can we help people who have difficulties in use a mobile app or a browser to order pizza? Is it also possible to make the ordering process more entertaining (or less boring)? With the increase in Internet of Things (IoT) \cite{Jenkins2015, ZMP007} devices entering our daily lives in the form of smart home devices such as plugs, thermometers and speakers people are now becoming more accustom to seeing and using these interactive devices \cite{Perera2015a}. In our work, we look at utilising this current trend to look at new ways in which pizza ordering can be accomplished outside of the normal ways to answer the 3 questions stated previously.

The increase in takeaway delivery in the last decade has shown massive market growth and is now worth an estimated GBP 4.2 billion as of February 2018 and is expected to continue to grow in the coming years. In 2017,   servings of pizza were ordered with roughly 38 per cent of those orders being online and with smartphone apps making up 16 per cent of the total orders, both of which are expected to be much higher as of 2019 \cite{Lavenant2018}. Our work will look at how we can make the ordering process more entertaining and explore areas where we are able to help users that struggle ordering food with smartphone applications or through the browser experience \cite{Tan2010}. An example target audience would be suffers of Parkinson disease whose dexterity and accuracy when using a smartphone or mouse on a personal computer leaves them unable to experience the convenience that these ordering systems provide.  As for the many other users of smartphone applications/browser as well as customers that order at the store level, we look at options that provide a more entertaining alternative to the repetitive process of ordering a pizza while keeping overall production cost relatively inexpensive.


In order to achieve our goals of providing an interactive, easy to use and entertaining system we  used a micro-controller with an RFID shield. Using RFID tags, the system can recognise pizza toppings or extras while still maintaining the simplicity and approachability. The device  encapsulated around a 3D printed ``pizza'', the design of which has been through prototyping and usability testing with a range of participants with varying levels of computer literacy. Testing with this range of participants allowed the final product to be approachable and easy to use by any age range that would want to make an order. The final stage of the prototyping was conducted on the premises of a pizzeria where customers would make their order using the system which, through real-world testing, provided feedback that catalysed discussions on future works. It also allowed us to look into how the user's emotions altered with the use of PizzaBox, whether it be positive or negative emotions such as joy or frustration. Care was taken with these tests as a change or deviation from a process that customers are accustomed to for processing their order can have an adverse effect on customer satisfaction as discussed in \cite{Dixon2009a} and thus an adverse effect in sales for the company that is utilising the new system that we have designed.

In Summary, through an high fidelity prototype, we examined  opportunities and challenges of using a physical object manipulation based technique to augment the food ordering process to create a novel experience for end users. Our prototype driven approach allowed us to gather  realistic and valuable insights from range of potential end users on various topics (healthcare, advertising, entertainment, lifestyle).

\section{Background}

This paper lies at the intersection of food ordering, tangible objects, tabletop designs \cite{Apted2009, Weiss2010}, and connected devices \cite{Okada2016, Nicenboim2018}. We also review several of these relevant  areas in order to borrow useful concepts as well as to differentiate our work from existing work.

\subsection{Food Ordering}
When we think of ordering a pizza, there are a plethora of ways in which we can achieve that. Dominoes are currently taking advantage of all aspects of available technology to give their customers the ability to order their favourite pizza wherever they are and with with whatever technology they have available. Apart from the more commonly known website and smartphone application ordering systems, Dominoes now provide the availability to order pizza over Google or Amazons digital assistants, Google Home and Amazon Alexa respectively \cite{Sciuto2018}, as well as branching into in-car systems and allow Ford vehicles with \textit{`Ford Sync'} to order a pizza while driving \cite{Dominos2018}. 

\begin{figure}
	
	\centering

	\subfloat[Customer starting the order.]{\label{figure:1}\includegraphics[width=60mm]{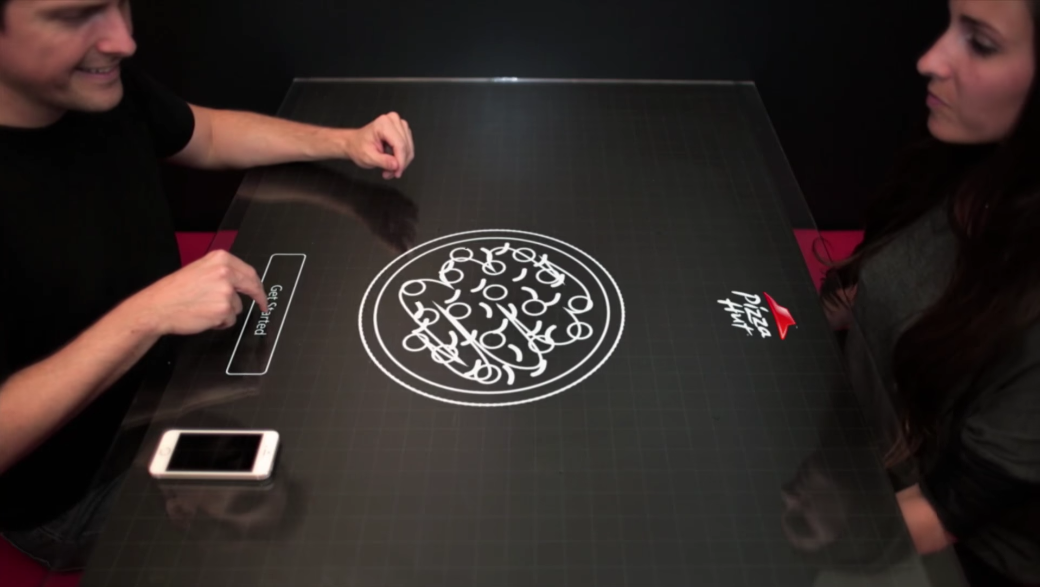}}
	\subfloat[System reacts by touch and dragging ingredients.]{\label{figure:2}\includegraphics[width=60mm]{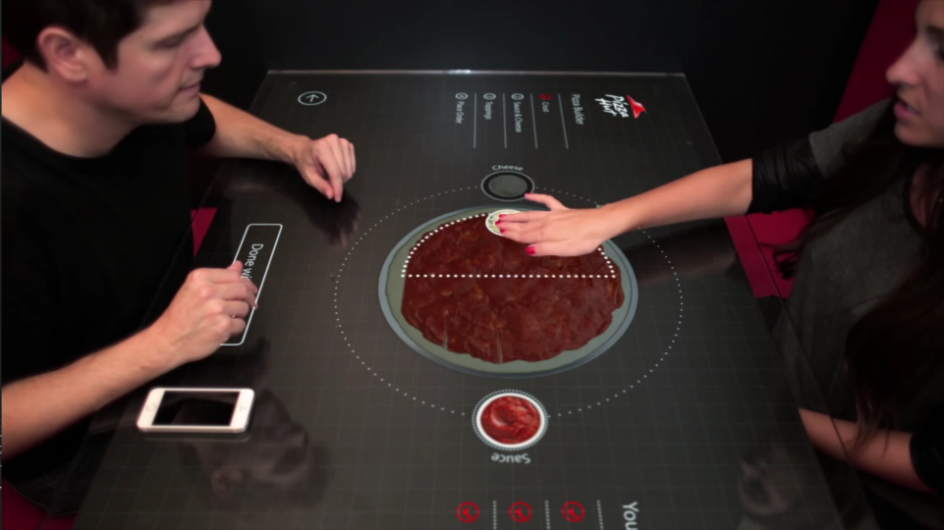}}
	\\
	\subfloat[The system allows for half and half pizzas depending on where the ingredient is placed.]{\label{figure:3}\includegraphics[width=60mm]{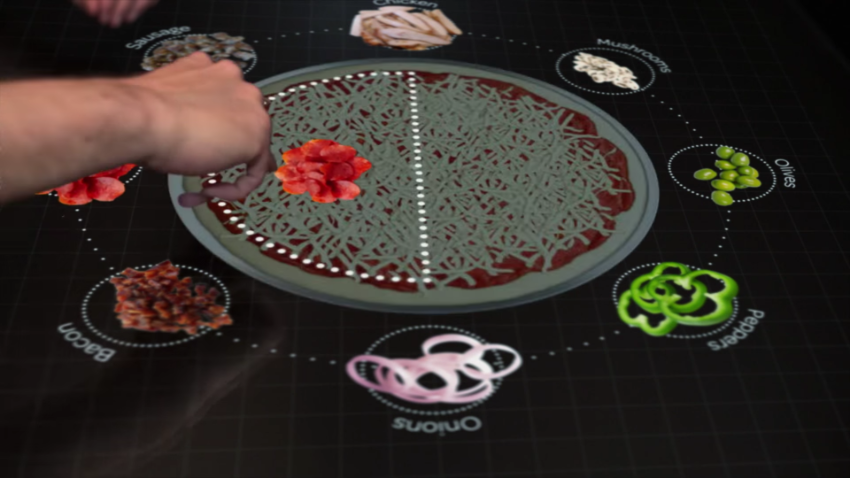}}
	\subfloat[Payment process is built into the tabletop system.]{\label{figure:4}\includegraphics[width=60mm]{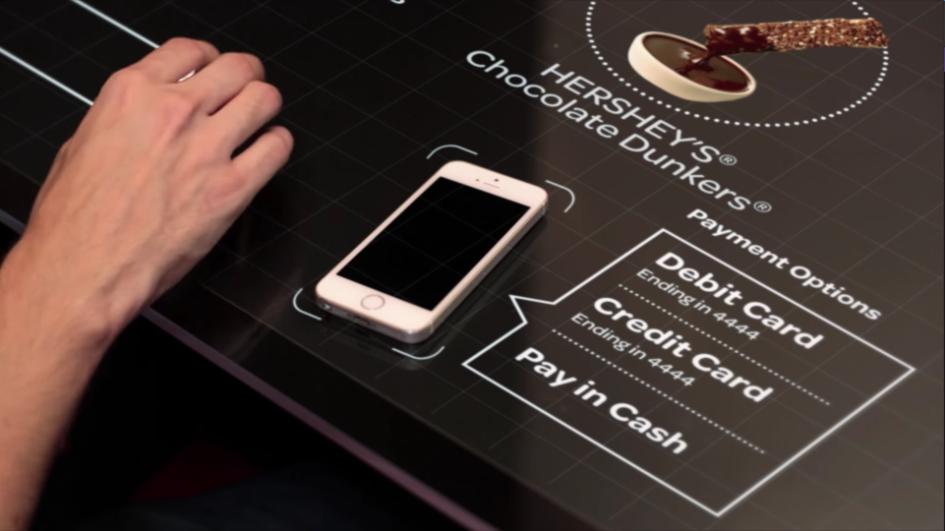}}
	\caption{PizzaHut's innovative tabletop ordering system. Screenshots taken from \cite{PizzaHut} \cite{Zolfasharifard2014}. This work is build upon techniques such as configuration switching presented in \cite{Ackad2010}. Tabletops are also used in other domains such as Museums to create playful environment \cite{Storz2014}.}
	\label{figur}
	\vspace{-10pt}
\end{figure}

Two of the most popular ways to order pizza are through smartphone applications and over a restaurants web site. Both of these platforms have been propelled forward with the decrease of cost and increased availability of PCs/Laptops and smartphones, each of which being able to provide aesthetically pleasing and intuitively easy to use food ordering systems. In 2015 Domino's noted that 77\% of all orders are now made online \cite{Jamieson2015} increasing their sales by 19\% \cite{Sheffield2015}. As illustrated in Figure \ref{figur}, restaurants are also working on new innovative ideas for ways to order food in the restaurants with PizzaHut revealing a tabletop concept for ordering pizza where the customer can drag ingredients onto a pizza and watch as the pizza is created in front of them \cite{Zolfasharifard2014}. The tabletop idea allows the customers to be able to order from their table creating a new social experience for the customers. A video \cite{Zolfasharifard2014} of the PizzaHut concept shows each customer interacting with the tabletop system which encourages new social cues for discussion based on what each customer would like on their pizza or vice versa. As the tabletop system also allows for each person to interact and add their own touch to the final order it provides a sense of involvement in the order process which isn't found when simply interacting with a pizzeria employee or through mobile or web ordering systems.

Furthermore, in 2016 fast food giant McDonalds introduced interactive food ordering systems into their restaurants. In a different way to PizzaHut's table-top system, McDonald's system consisted of touch screen devices which allowed customers to order their food instead of waiting in line to speak to a cashier. Interestingly, according to an article by The New York Times \cite{STROM2017} it was found that families and groups have been the biggest users of this system. This suggests the system not only introduces new, faster ways of ordering food for groups but also provides new social experiences based around the touch screen as each member of the group is able to interact with the touch screen to personalise their order.

We examined several inventions and patent applications  that used the food ordering sector as inspiration and how they aim to improve it with their inventions. The first patent application submitted in 1988 by inventors Michael and Daniel Dubno \cite{Dubno1982} looks to incorporate video games into the tables that  customers sit at in a restaurant and provide them with entertainment while they wait as well as given them the ability to order their food through the same computer system. Every table incorporates at least one monitor and the necessary gameplay equipment that allows for the engagement of two customers or any individual to play a solitary game. Each of the computer systems in the table then connects to a central server which is connected via an output displaying the orders of customers to the kitchen staff. The aims of this invention are to help alleviate the human error aspect that comes with traditional ordering techniques of writing down customer orders on mediums like notepads, communicating that order verbally or visually showing the notes to kitchen staff who then complete that order. Aside from the advantages to the ordering system, it also provides numerous advantages to the customer giving them a trigger to engage in social interaction and to remove the irritation that might be induced by delays and setbacks in food service.

The second patent  was an interactive food and drink ordering system invented by Riddiford et al. \cite{Riddiford2007}. The invention aims to remove the stress on a waiter or waitress taking orders during busy periods and by doing so lower the risk of human error while taking orders. The invention looks to decrease the amount of effort and cost that comes with implementing an embedded touch screen monitor in the tables and instead use a computer controlled projector that is mounted above a table or other surface that can project a display of the menu. The customer then selects the items they wish to order by operating a device that is connected to the computer above them via Bluetooth. The items that have been ordered will then be displayed on a monitor to the kitchen staff to fulfil. The invention also gives the option for the customer to access a \textit{`chef-cam'} to provide them an insight into how their food is prepared and cooked which gives the customers some entertainment or distraction while they wait for their food to arrive.

Both of these patents follow a few similar ideas that are applicable to our project. Firstly, both inventions look to take the pressure off a waitress or waiter/chef which is something that we hope to achieve with our project should it be set in a restaurant environment, especially an independent pizzeria that is run with minimal staff. Secondly, both inventions provide a form of entertainment for the customers giving them something to distract themselves while waiting for their food or providing social cues to enhance their experience at the restaurant.


\subsection{Surfaces/Tabletop Designs}
The \textit{TangiSense} \cite{Kubicki2015} is a system in which the users are able to associate information with behaviours to manipulate tangible objects. The system continued to express the strength of tracking or identifying objects using RFID technology and expressing that tracking through the use of LEDs and sound. However, this system restrains the use outside of a tabletop environment as it uses multiple RFID antennas over an area of 1m x 1m to be able to read multiple items on the table top. The \textit{Sensetable} \cite{Patten} is a tabletop wireless tracking platform for tangible user interfaces that utilises electromagnets to track items on a tabletop display surface. The researchers on this project aimed to provide accurate and low-latency tracking for 6 - 10 objects and to be able to make physical changes to modify the tracked objects. Although the technology in this paper is not applicable for our project, the idea of tangible objects is enforced with the feedback from the test participants saying that the ability to use and manipulate the objects is `very exciting' and being able to gain intuition on what was happening quickly.

\subsection{Tangible Objects}
The Ambient Birdhouse research project that was conducted at the Queensland University of Technology worked with the idea of using IoT devices to connect users to their outside environment by helping to provide enlightenment of the birds in the area through playfulness and entertainment. \cite{Soro2018} Users are prompted to engage with the system at pre-determined intervals as the birdhouse starts playing a video of a bird in its natural environment. As well as this it allowed interaction with the system through the use of cards that would start guessing games, videos or video uploads. The cards communicated with a Raspberry Pi Model 3 that controlled an RFID shield to identify what card was being presented to the system.

The Empowering Occupational Therapists \cite{Moraiti2015} project brought forward the idea of using smart soft objects to help caregivers create tailor-made assistive solutions. Using `The Skweezee system' which utilises an Arduino board and Multiplexer with some sensing electrodes to be able to hack everyday objects to re-purpose them for therapy. This system highlighted the ability of the Arduino boards strength when working in the space of interactivity between objects.

The smart objects and forgetfulness MessageBag \cite{Farion} prototype was designed and created with the idea of combining technology and everyday items to alleviate the stress of forgetting objects such as keys or a wallet. The project embraces the idea of using ubiquitous computing to embed technology into a bag which can communicate with the users visually using LED lights to easily show what is and isn't in the bag, showing the user that they've forgotten to pack something. The prototype utilises LEDs, a Teensy Board and an ID-12 RFID reader which reads tags with unique ids when placed into the bag. When an RFID tag is read the corresponding LED that is assigned to the unique ID of the tag will turn off providing the user with a visual cue that the item has been added to the bag as well as what they still need to pack. This project highlights the importance of visual and audio cues when users are looking for feedback to an action, the LEDs provide this and are inexpensive to incorporate. Both  tabletops and tangible objects\cite{Khandelwal2007, Girouard2007} are used to facilitate effective pre-K math education.

\begin{figure}[t]
	\centering
	\includegraphics[scale=0.53]{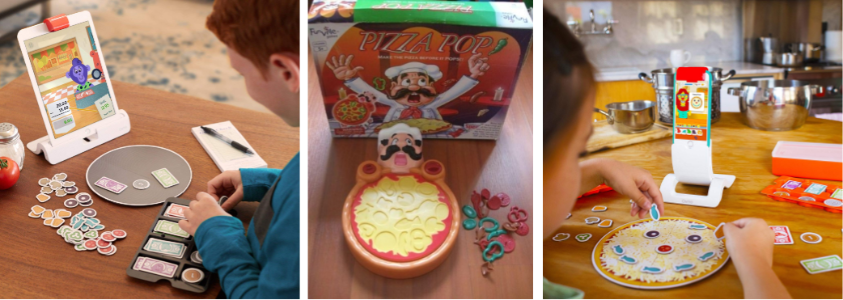}
	\caption[FunVille-Games Pizza Pop]{The game board and ingredient elements used to play the game \cite{PizzaPop}. Clear design element similarities here helped reinforce the design aspects of our work.}
	\label{fig:childreninteractivegamepizzapop}
	\vspace{-10pt}
\end{figure}

Commercially there are many games and toys that are aimed at a younger audience that is based around the process of ordering pizzas. One of our aim is to provide an entertaining option with the order process that goes along with ordering pizzas so naturally looking at games and toys was our first go to for aesthetic and ergonomic designs. FunVille-Games released a relatively unknown interactive game called \textit{`Pizza Pop'}. As shown in Figure \ref{fig:childreninteractivegamepizzapop}, the premise of the game is that the player places 3D ingredients on to a pizza base in the allocated places before the pizza base erupts, throwing all the ingredients off the base. This was found by a colleague and presented to us after we had finished our prototype designs. This board game was useful to reinforce some design elements and concepts that we have used in the early prototypes as they can also be seen in the interactive game.

\textit{\textbf{Summary:}} As discussed before, we evaluated a collection of patents as well as innovative ideas by large food  companies that try to make the food ordering process more entertaining and streamline, taking away the norms of waitress/waiter interaction and replacing them with interactive computer systems that can process food or drink orders as well as take the payment, giving the customer a new experience as well as new social interactions \cite{Ng2016}. We also notice the strength of Arduino or likewise micro-controllers in the space of ubiquitous computing with both the Empowering Occupational Therapist \cite{Moraiti2015} and MessageBag prototype \cite{Farion} utilising them to great advantage. Another technology, RFID, has been utilised often when projects require object detection or tracking which for our own project is a key aspect. Although each project uses this technology in a different way it provided inspiration for our design \cite{DeFreitas2016}. Our design also inspired by TANGerINE \cite{Baraldi2007} which demonstrate how to utilise tangible objects to facilitate interactive natural environment.

\section{Designing Connected PizzaBox}

\subsection{Methodology}

\begin{figure}
	\centering
	\includegraphics[scale=0.57]{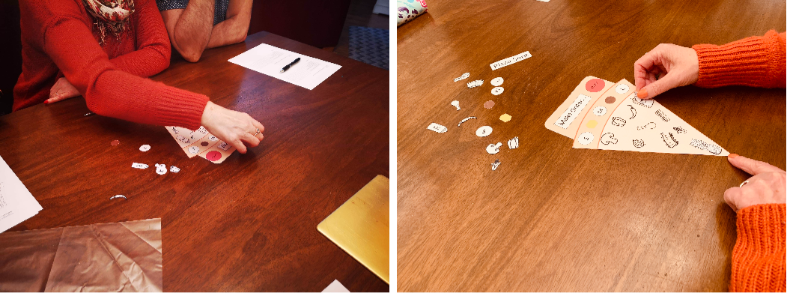}
	\caption[Participants engaging with a prototype]{Participants engaging with a prototype in round one. Use of the prototype was unguided to understand how intuitive and easy to use the prototype was.}
	\label{fig:img0957}
\end{figure}

\begin{figure}[t!]
	
	\centering	
	\subfloat[Participant adding an ingredient to the  prototype.]{\label{figur:1}\includegraphics[width=78mm]{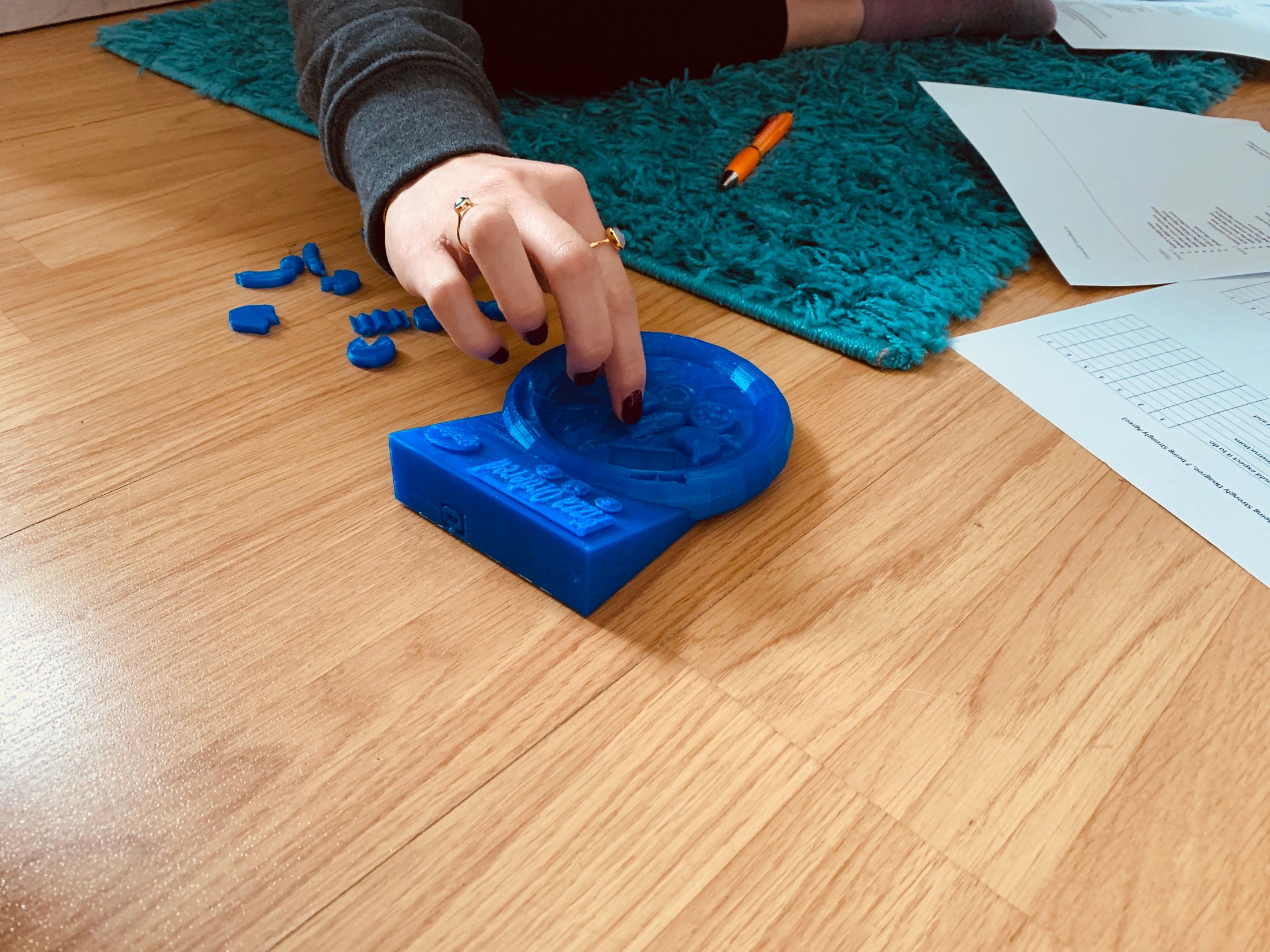}}
	\subfloat[Participant completing a task by pressing the order button.]{ \includegraphics[width=78mm]{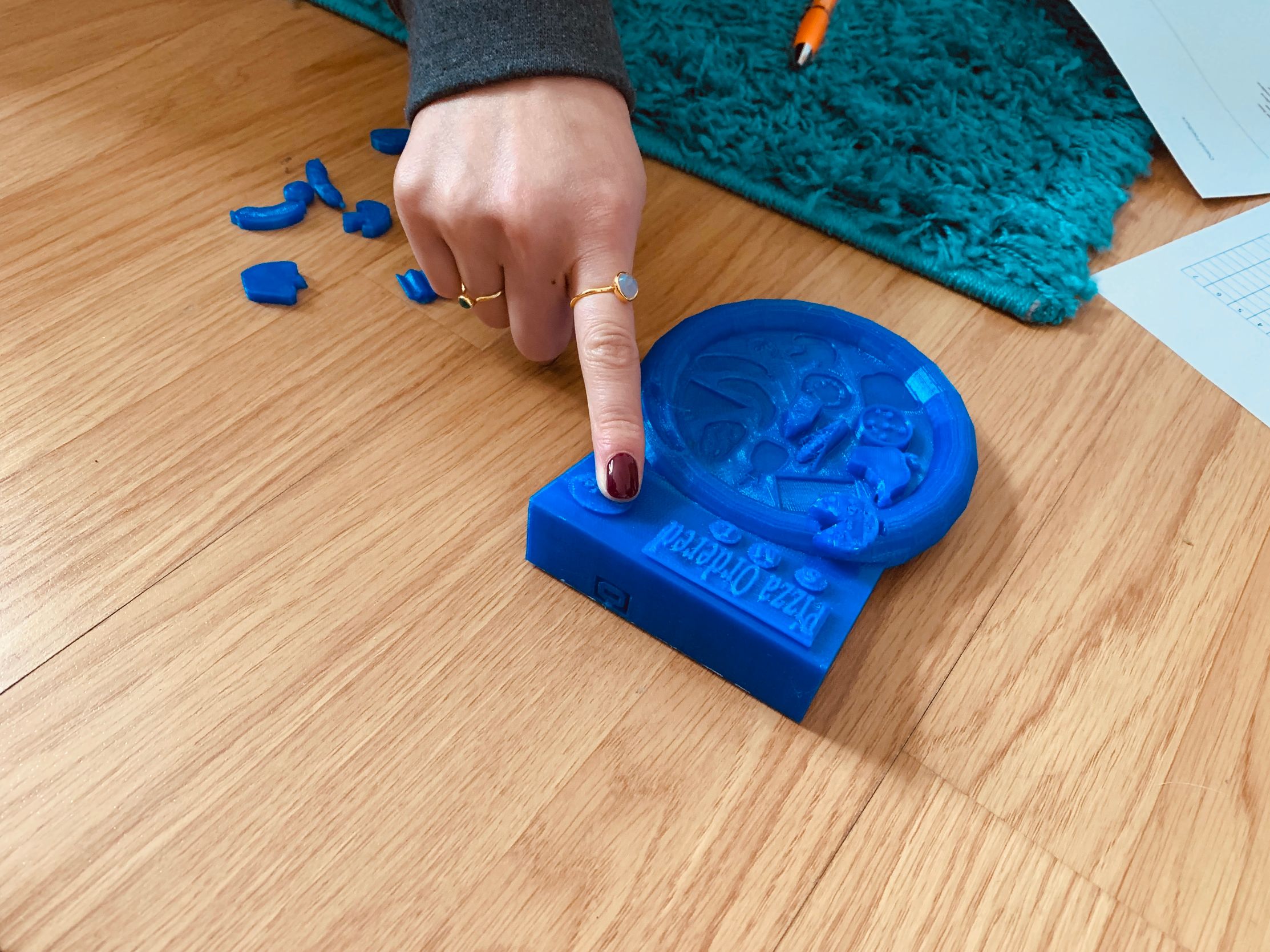}}
	\\
	\subfloat[Participant adding an ingredient to the prototype.]{ \includegraphics[width=78mm]{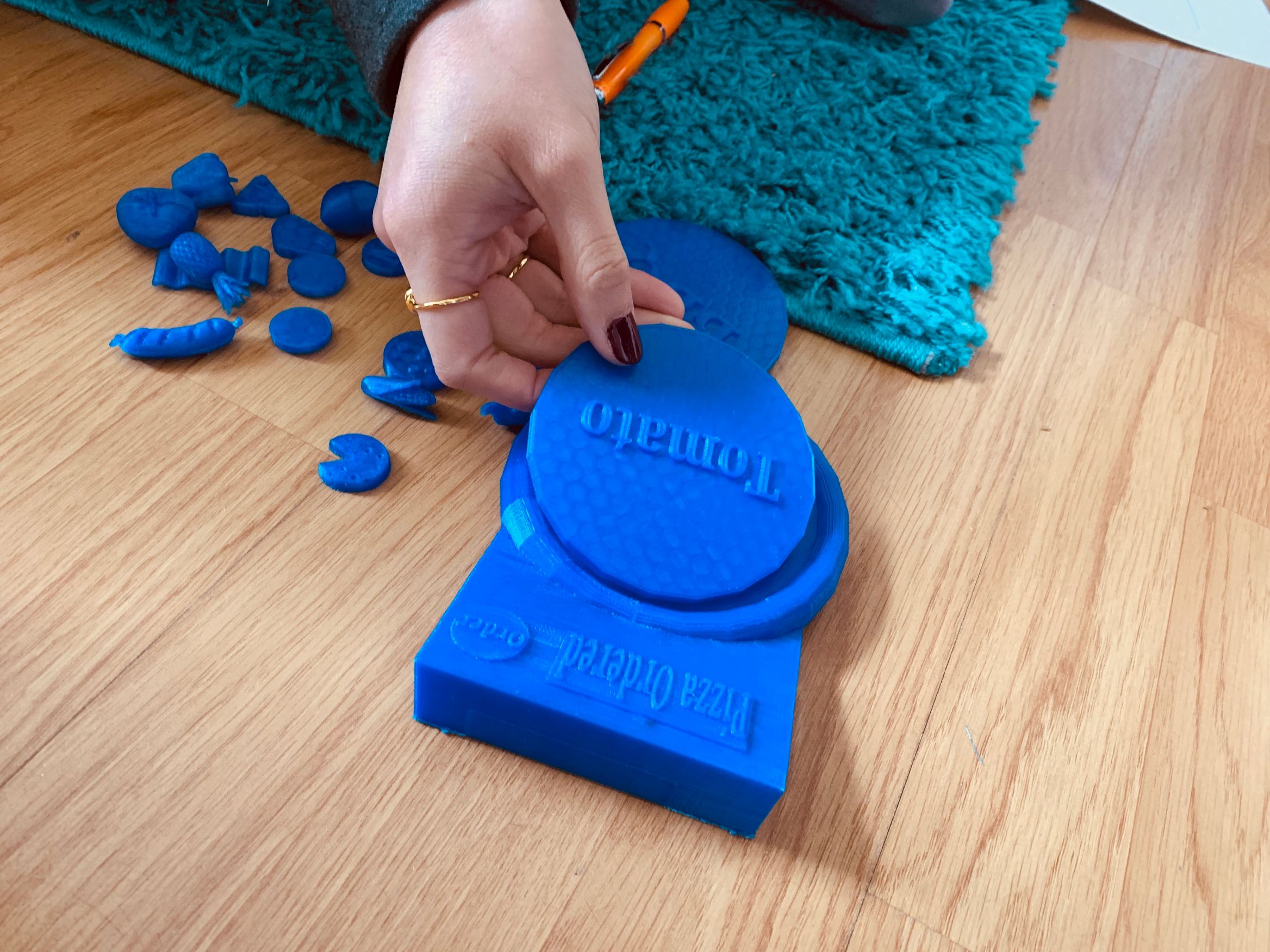}}
	\subfloat[Participants working together to complete a task.]{ \includegraphics[width=78mm]{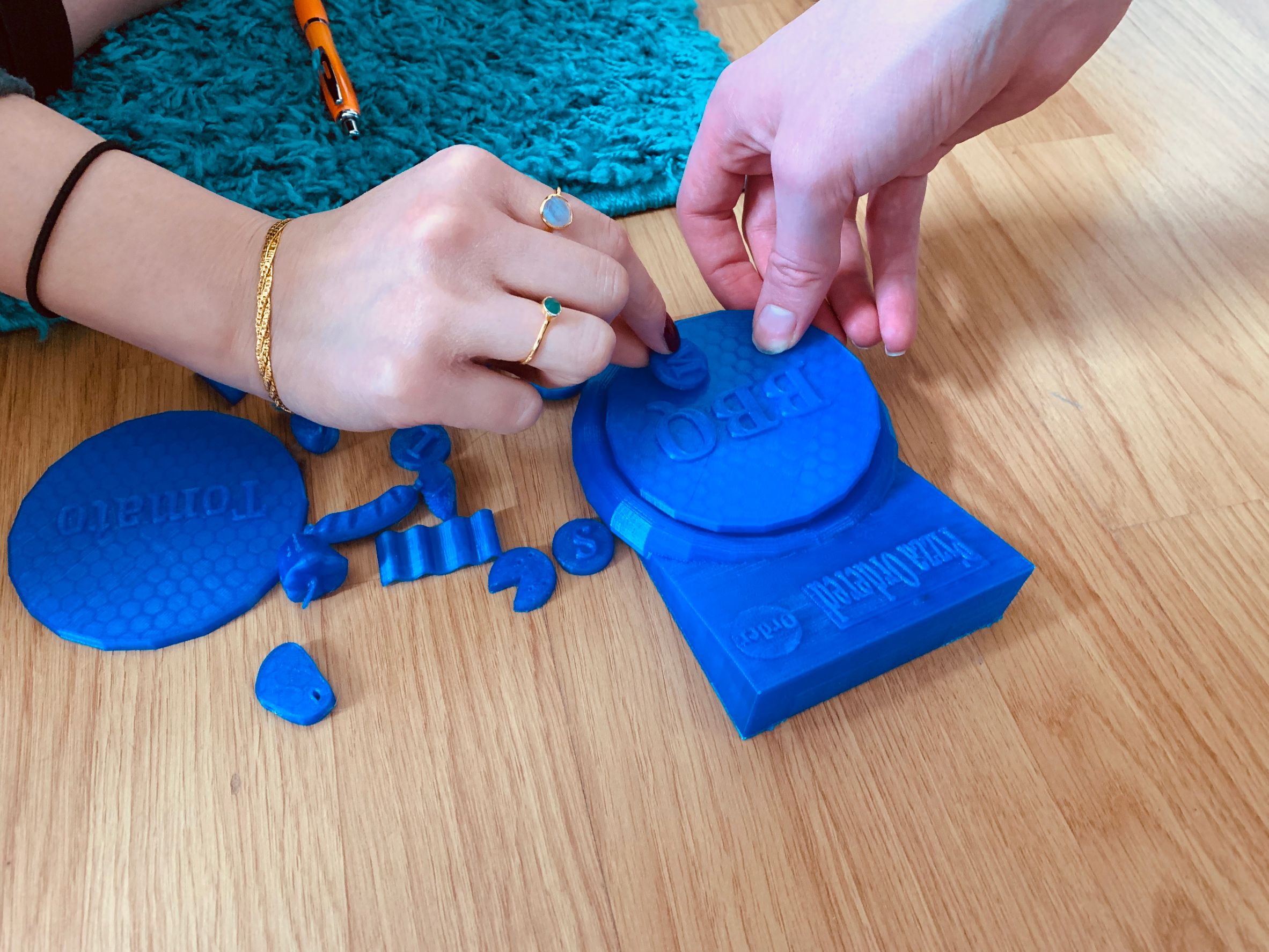}}

	\caption{Participants engaging with the prototype during the study 2 medium fidelity tests.}
	\label{Figstudy2}	
	
	\vspace{-10pt}
\end{figure}

\begin{figure}[!t]
	\centering
	\includegraphics[scale=0.18]{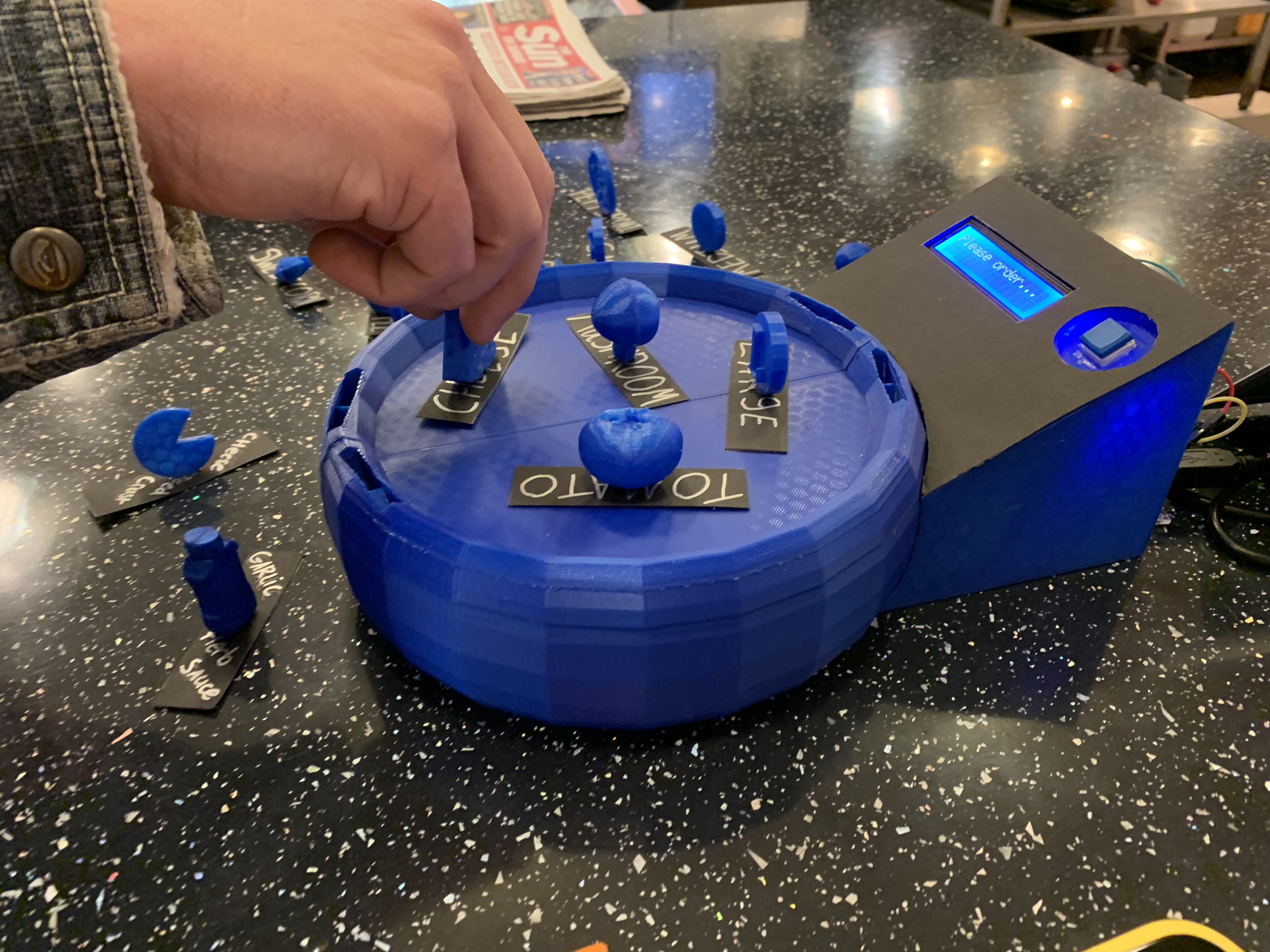}
	\caption[final evaluation participant interacting with PizzaBox]{Participant  placing 3D printed ingredients onto the base of the PizzaBox before pressing the order button.}
	\label{fig:img1157}
	\vspace{-10pt}
	
\end{figure}

In high-level, we conducted three rounds of focus groups studies that involved co-design \cite{Mazalek2009} activities. Study 1 utilized low prototyping techniques (Figure \ref{fig:img0957}) where study 2 utilised medium-fidelity (Figure \ref{Figstudy2}). In study 3, we tested our high-fidelity prototype (Figure \ref{fig:img1157}). After gaining approval from University Ethics Committee, we used an open local university wide mailing list to recruit 5 groups of 3 participants. 2 groups from a STEM subject, and 2 groups from a non-STEM subject. The last group consisted of adults out of education and in full-time work. Due to ethical reasons, participation was not open to anyone under the age of 18. We forwarded the information sheet approved by the Ethics Committee to the participants, which explained the study in more detail and their rights. Each focus group was recorded, both audio and video for further analysis of feedback and how the prototype was used. No video was taken of the participants faces only a top-down view of the prototypes and their hands using the prototype to protect anonymity and respect privacy.

\subsection{Refining and Curated of Alternatives Designs: Study 1}

\subsubsection{Scenarios}
\label{sec:Scenarios}

In this study, each group was provided with set tasks to complete while using the prototype. Each task contained 6 to 8 steps and aimed to utilise as many of the features of the prototype as possible. 


\begin{itemize}
	\item Task 1: Could you order me a tomato base pizza medium onion mushroom beef / cheese stuffed crust?
	\item Task 2: Could you order me BBQ base large pineapple sweetcorn and chicken?
	\item Task 3: Could you order me tomato base medium pineapple chilly and pepper?
	\item Task 4: Could you order a pizza of your own?
\end{itemize}


For example, in order to complete task 1, the users would need to complete following steps: 1) add the pizza base 2) select the correct size 3) add onion ingredient 4) add mushroom ingredient 5) add beef ingredient 6) select stuffed crust option 7) press order button. We used the same tasks across all participants to maintain consistency and to give us the ability to look back at the video recordings and analyse how each group of participants differed in their use of the prototype. The last task (i.e., Task 4) allowed the user to put themselves in a real-life situation where they were free to order whatever pizza they desired at the time. Doing this allowed the users the opportunity to design a pizza how they would out of the testing space which gave us the opportunity to gain more valuable insight into each prototype. Firstly, to reiterate one of our goals for the project, we aim for the system to be easy to understand and learn without the need of expert help instruction or manual so this final scenario of giving the participants the free will to order what they wanted, allowed us to gain an understanding of whether the participants fully understood how to design a pizza and make the order. This last scenario also shone a light on clear restrictions or limitations of each prototype as participants often had their favourite pizza combination which sometimes included an ingredient, dough option or crust option that the prototype was unable to provide.

\subsubsection{Study 1: Low Fidelity Prototyping}

Each group received a questionnaire to fill in and forms for leaving feedback on each of the prototypes that they will be shown. For each group, we first explained in more detail what our project entailed and then asked them to work together and sketch out their own idea for a prototype to fulfil the goal of our project. The idea behind this was to find common elements that each of the groups came up with and whether those common elements appeared in our own prototypes or the opposite if they didn't appear in our prototypes then we could consider a possible design change.

\begin{table}[t!]
	\caption{Highlighted Feedback from individual participant on each individual prototype.}
	\label{TblParticipants}
	\centering
	\footnotesize 
	\begin{tabular}{p{0.2cm} p{1.2cm}  p{2.4cm} p{2.4cm} p{2.4cm} p{2.4cm} p{2.4cm}}
		
		\begin{sideways}Group ID    \end{sideways}
		
		& 
		{\scriptsize \begin{sideways} \begin{tabular}[c]{@{}l@{}}Age Group \\ (Number of \\Smartphone App) \\ {[}Hours Spent \\on Phone/PC] \end{tabular}  \end{sideways}}	 
		& 
		\begin{sideways} \begin{tabular}[c]{@{}l@{}} Notable feedback \\(Prototype 1)    \end{tabular} \end{sideways}	                                                                          & 
		\begin{sideways} \begin{tabular}[c]{@{}l@{}}  Notable feedback \\ (Prototype 2) \end{tabular}                                            \end{sideways}	                                                                & 
		\begin{sideways} \begin{tabular}[c]{@{}l@{}}  Notable feedback \\ (Prototype3) \end{tabular}                                               \end{sideways}	 	                                                          & 
		\begin{sideways} \begin{tabular}[c]{@{}l@{}}  Notable feedback \\ (Prototype 4)  \end{tabular}                                      \end{sideways}	                                               & Overall Feedback                                                                                                                 \\ \hline
		1            & \begin{tabular}[c]{@{}l@{}}18-25 \\(41-50+)\\ {[}7-10+] \end{tabular}                    & Stuffed crust options ambiguous or   confusing.   Missing option for half and half.  & No option to double up on   ingredients.   Missing gluten free or no cheese   option.    & Confusion about whether all slots need   filling to complete pizza.   Visually unappealing. & No option to double up the ingredients   or to remove/add cheese.     & No control over cheese topping.   Alteration to pizza size selection   (Slider).\\ \hline
		2            & \begin{tabular}[c]{@{}l@{}}56-65 \\(21-30)\\ {[}0-4] \end{tabular}                     & Brown/Yellow icon were not clear.   No option to double up or decline cheese.      & Looks as though only ordering a single slice of pizza.                                                                                    & Size selection should be buttons to press.                                                                                              & No double up option.   No option to decline cheese.                   & No control over cheese toppings.   Stuffed crested functions unclear.           \\   \hline
		3            & \begin{tabular}[c]{@{}l@{}}18-25 \\(11-20/\\21-30)\\ {[}7-10] \end{tabular}                     & Needs instructions.   No option for dietary requirements.                          & Limited topping options.                                                                                                                  & More organised than others.                                                                                                             & Easier to understand how to place   the ingredients and order the pizza.  & Lack of allergy options.   All engaging to use.                                \\    \hline
		4            & \begin{tabular}[c]{@{}l@{}}18-25 \\(21-30/50+)\\ {[}7-10] \end{tabular}                   & Needed description of stuffed crust options.                                                                                    & Not sure whether ordering full pizza or just a slice.   Buttons easier to understand for sizes.  & Half and half limitations and double ingredients.    Not as fun as seeing the ingredients. & Layout of topping better than first prototype, more structured.                                                    & Needs clearer indication when pizza order is sent.   All needs to be bigger in size.  \\  \hline
		5            & \begin{tabular}[c]{@{}l@{}}18-25\\(0-10/\\11-20/50+)\\ {[}7-10+] \end{tabular}                    & Confusing where to put the size   options as they are same size as order button.   & Good user interface.                                                                                                                      & The amount of ingredients is limited   to 6 after choosing sauce and size.                 & Needs feedback for what size has been   selected.                     & Limitation on half and half.                                                                                                   \\ \hline
	\end{tabular}
	
\end{table}

After designing their own prototype, we would then move on to show each of our own 4 prototypes and without explanation of how the prototype was intended to be used, we asked the participants complete tasks laid out in the scenario that we discussed in the previous subsection. Throughout each prototype, we constantly engaged with the participants to keep them thinking aloud and expressing their thoughts and opinions, with the most notable and reoccurring feedback shown in Table \ref{TblParticipants}. Using the think-aloud method encouraged the participants to point out features that the prototype was lacking i.e. if they regularly ordered a \textit{`half and half'} pizza but the prototype lacked the capability to order that style of pizza, or if the feature was there but not intuitive to use.


\subsubsection{Participant Recruitment}
\label{sec:ParticipantRecruitment}

This section describes the participants with data taken from the questionnaire that was filled in upon arrival of the first stage of testing and will give an overview of the technical capabilities by a very simple assessment of how many applications each participant has installed on their phone and use of technology on a day to day basis. All participants were aware of current pizza ordering systems so were able to use that experience to compare and contrast the prototypes with technology that is already available and to help visualise the prototype. These groups were used for both the low-fidelity and medium-fidelity testing.

\textbf{{Group 1: Participants P1, P2, P3}}: The first group of participants were all from a STEM subject, specifically Computer Science and in their last year. Two of the participants were male, P1 and P2 both aged 18 - 25 years. P3 is the only female participant in the group, aged 18 -25 years also. All participants had a high level of technical skill and experience, all of which describe themselves as expert users of IoT and digital technologies. Each participant had 50+ applications on their mobile phones and each spends 7 to 10+ hours per day interacting with either a smartphone or PC/laptop. 

\textbf{{Group 2: Participants P4, P5, P6}}: Group 2 participants consisted of adults out of education and in a working environment. Two of the participants were female, P4 aged 56 - 65 and P5 aged 18 - 25. P6 being the only male in the group, aged 56 - 65. Participants P4 and P6 have a low degree of computer literacy, using their mobile devices for basic needs and only having 20 - 31 applications installed. General use of digital technologies for these participants was also low with 4 -7 hours of use per day and the majority of that use being at work to fulfil basic work tasks. P5 had more experience with digital technologies but stayed consistent with the other participants in terms of the number of applications installed on their smartphone device and time spent using their PC/Laptop.

\textbf{{Group 3: Participants P7, P8, P9}}: This group consisted of two females P7 and P8, and one male P9, with all participants aged 18 - 25. Each participant studied humanities subjects in their 3rd year of university. Each participant, although not studying a technical degree, described themselves as technically minded as their hobbies and interests outside of their humanity studies involved technical aspects. Each participant had different levels of applications installed on their mobile devices from 11 - 20 up to 50+ (This was later found out to be due to the age one of the participants mobile device was, restricting their availability of apps) but the amount of time they spent using a PC/Laptop/Mobile stayed similar around 7 - 10+ hours per day.

\textbf{{Group 4: Participants P10, P11, P12}}: Group 4 participants consisted wholly of female participants, aged 18 - 25. All the participants were in their last year of studying computer science and all demonstrated a high level of computer literacy. One participant, P10, also extended her interest in technology beyond her degree and into her hobbies. P10 and P11 both have 50+ applications installed on their mobile devices with P12 having 21 -30. All participants spent the same amount of hours per day using their PC/Laptop/Mobiles at 7 - 10 hours.

\textbf{{Group 5: Participants P13, P14, P15}}: Our final group consisted of another set of students studying humanities subjects. The group consisted of two female participants, P13 and P14 and 1 male participant P15. All participants in this group were aged 18 - 25. Participants P13 and P14 both described themselves as having a distinct lack of interest in technology and used their mobile phones for just social media applications with a total of 11 - 20 applications. P13 spending 7 - 10 hours per day using technology and P14 spending 7 - 10 hours a day using technology. P15 describes themselves as someone who dabbles in technology but only at a high level and has numerous games and applications on their mobile device at 50+ and spending 10+ hours per day using mobile technologies.

\subsubsection{Data Analysis}
From the feedback gathered from the initial tests, we could see there were clear drawbacks for each prototype but the most highlighted negative feedback for each prototype overall came down to three things. 

\begin{itemize}
	\item The lack of control over allergens or special dietary requirements. Furthermore on this topic was the confusion when it came to the cheese topping. Each prototype was developed with the idea that each pizza would automatically have a cheese topping as per a generic pizza. This presumption became a clear drawback with many of the users experiencing confusion to whether there was cheese on top of the pizza or if they could remove the cheese due to special dietary requirements. 
	
	\item The lack of a \textit{`half and half'} feature. Over the years big pizza companies such as Domino's, Pizza Hut and Papa Johns have all been creating new ways for customers to enjoy their food and one of the most popular ways is the \textit{`half and half'} which allows two different pizzas on each half of one pizza. Our prototypes lacked this feature and were talked about often during the tests and was one of the most common bits of feedback we received. 
	
	\item Although we were not necessarily looking for this during the initial testing as we were concentrating more on design, many of the participants (especially from a STEM degree) commentated on the system design e.g. what if they placed more than one size on the prototype and how the system would react to that. These conversations also helped with the early design of the system and how we can build on previous idea's to help create a better user experience with our end product.
\end{itemize}

\subsection{Refining and Curated of Alternatives Designs: Study 2}

\subsubsection{Scenario}
The scenarios used for the medium fidelity tests were kept the same as the low fidelity tests. Please refer to Section \ref{sec:Scenarios} for further explanation of the scenarios. For the medium fidelity prototype tests, the `free will' scenario also gave the participants the opportunity to use other 3D models that had not been utilised in the previous three tasks so they can provide a better quality of feedback when looking at the aesthetics of the model.

\subsubsection{Study 2: Medium Fidelity Prototyping}
After the completion of the low fidelity stage of prototyping and analysis of the results from the Likert scale questions as well as the feedback received for each of the four prototypes, we decided on two prototypes that were to be taken forward and turned into a 3D model to then be used in the medium fidelity tests. Prototype 2 and 4 had on average higher results with each of them receiving good strong feedback.

From the data collected from study 1 slight alterations were made to the prototypes that we presented to the participants in study 2. The first alteration was the addition to control the cheese topping as this was the main bit of negative feedback we received from the first study. Second was to make the feature of adding stuffed crest options easier to understand by using 3D printed ingredients that slot directly into the crust of the pizza prototype. We hoped this would remove the ambiguity of the original design of having yellow or brown hexagon shapes to represent cheese and sausage stuffed crust option that just slotted into the crest section of the prototype.

For the medium fidelity testing stage, we still encouraged participants to give feedback on the design but our main focus of this iteration of tests was to look at the ergonomics and aesthetics of the prototype. We asked the same Likert scale questions as the low fidelity tests as well as added questions that related to the size of the prototype overall, the size of ingredient parts, the logical layout of the product and how easy components were to recognise. As the idea behind the product was to make ordering a pizza fun and unique experience, having small or awkward components that caused frustration, for example, would negatively affect the experience for the user. 

The medium fidelity prototyping tests were conducted under similar conditions as the low fidelity tests. We were unable to use a camera to record the test due to faults with the camera but meaningful comments made by the participants were noted and photos of interaction with the prototype were taken for further analysis of participant interaction with the prototype. Each group was shown the prototype and asked to carry out a series of tasks as per the low fidelity prototype tests. All participants were asked to think about the aesthetic elements of the components when completing the tasks and to give feedback verbally or to write down the feedback when filling in the Likert scale survey questions that had been provided to them. We emphasised with the participants how verbally providing feedback was a good idea and asked questions to encourage it as we felt having each user try to remember the feedback until they finished the tasks would cause them to forget important feedback.

\subsubsection{Participant Recruitment}

For the second stage (study 2) of prototyping, we maintained the same participants that we recruited for testing stage 1 testing. Refer to \ref{sec:ParticipantRecruitment} on how we recruited these participants and a description of each participant in their groups.

\subsubsection{Data Analysis}
From the feedback gathered from this stage of tests, we were able to conclude that certain aspects of the design needed further alteration when designing the final product. As discussed earlier, we asked participants to think about the sizes of components, how easily recognisable ingredients were and how logically set out the prototype was. We received clear feedback on all these topics and in general, the feedback we received when looking at the size of the product, its components and the ingredient elements for both prototypes showed us that overall the product is too small and caused frustration when trying to pick and up and place the ingredients. It also showed us that the design of some of the ingredient designs proved to be ambiguous and explanation was sometimes needed on which design was the corresponding ingredient. These were important feedback and alterations to the design will be made to elevate these concerns from the participants.

From the Likert scale questions that we provided each participant, we noticed a trend of very similar results when comparing the same questions that we asked participants in the low fidelity tests with very slight increases to the values of how fun they found the experience and how easy they found the prototype to use. This could suggest that the better experience the individual had was due to the fact that they were able to interact with the 3D model instead of trying to visualise the product from a paper prototype in the low fidelity tests.

\subsection{Prototyping PizzaBox}




The project is headed by the idea that the PizzaBox will become a product that a family or restaurant can buy at a relatively low cost. The hardware that we have chosen to use in this project attempts to remain true to the low-cost ethos. Arduino Uno Wifi Rev 2 \cite{Arduno} was an ideal component for our project. The Arduino Uno Wifi Microcontroller allowed us to add on modules which were used to give visual feedback to the user via an LCD Screen as well as giving us the functionality of being able to read data from an RFID tag via an RFID reader. The Arduino Uno Wifi also incorporates a Wifi module which was ideal for connecting to the internet and sending the users orders via email. We originally intended to use an Adafruit PN532 NFC/RFID controller breakout board \cite{Adafruit2019}. This RFID board was relatively cheap  but unfortunately, when trying to work with the shield we discovered that there was a hardware limitation on the number of RFID tags the shield could read at any one time, the limitation being 2 tags which made it  unsuitable. 

Alternatively, we used SparkFun Simultaneous RFID Tag Reader \cite{SparkSRTR} which is a simultaneous RFID tag reader (SRTR). The reader uses the M6E-NANO chip  which can read up to 150 tags per second and have a variable range depending on antenna and power input. This was ideal for the project as we could scale down the range of the antenna to just the base of the product as to not pick up data from nearby tags and keep overall power consumption to a minimum.  To be able to work with this module we also started using the GEN2 UHF RFID Tags. These tags have more than enough memory for writing ingredient details to and are thin enough to be attached to the ingredients without any interference to the design of them. Adafruit Standard 16 x 2 LCD Display \cite{LCDdisplay} was used as it allowed for enough character space to give sufficient feedback to the user as well as having a bright back light meaning the text is easily read. We also used the I2C/SPI Character LCD backpack \cite{Adafruit2019a}  as it not only allows the use of the LCD through just 2 pins (excluding 5v and Ground pins) but it also allowed us to be able to have the LCD free from the Arduino board so we could place it in an optimal position for ease of use.


%

\section{Evaluation}

\subsection{Participant Recruitment}

The evaluation (study 3) was conducted in a local independent pizzeria and so participants were recruited upon visitation of the pizzeria. As participants were customers of the store, the demographics of each customer varied slightly with the majority of participants being 40 years of age plus with a mix of genders. Participants were asked if they would like to take part in the test and if they agreed a simple explanation of what the project entailed was given as well as explaining that any information shared with us remains anonymous and no data could be linked back to the participant. In this paper, we refer to participants as \textcolor{black}{C1 (Male, 21 - 30 years old), C2 (Female, 21 - 30 years old), C3(Male, 41 - 50 years old), C4(Female, 41 - 50 years old), C5(Male, 41 - 50 years old) .} For ethical reasons, participants under the age of 18 were not asked to participate unless accompanied by their parent or guardian. Participants were then asked questions from a question pool that we felt were applicable to the individual but also gave us full coverage of the data we wished to collect. As we were based in a local pizzeria the owner and sole worker of the pizzeria also agreed to participate in the evaluation test to be able to give a different perspective. \textcolor{black}{In the following sections, we refer to the pizzeria owner as P1 (Female, 41 - 50 years old)}. Our objective was to gather qualitative data in order to extract useful insights.

\subsection{Scenarios}

The scenarios were not pre-planned like the previous low and medium fidelity testing but were based on what the customer had recently ordered from the pizzeria. Each participant that was willing to take part in using the PizzaBox and answering questions based on their experience was simply asked to order the same pizza or similar based on what ingredients we had available, that they ordered with the pizzeria employee. This was agreed to be the best way to evaluate the PizzaBox as it gave the customer the ability to easily compare their experiences of ordering face to face with the pizza maker and then with the PizzaBox. It also gave an easy introduction into how using the PizzaBox could affect their choice of food order as they're able to see the ingredients they ordered making questions regarding eating habits easier to think about for the participant.


\section{Discussion}

We followed Miles \cite{Miles2013} framework to conduct the qualitative analysis. Further, for data reduction phase, we use Richards \cite{Richards} three tier coding technique (i.e., descriptive coding, topic coding, and analytic coding). The thematic areas we found by analysing the data as follows:


\subsection{Usability and End User Engagement}
This theme takes a deeper look at what the customers and pizzeria owner believe what age group of people that the PizzaBox system would have the highest appeal. We also discuss whether the PizzaBox would be utilised more by an individual user for by a group of users and why. Overall, most customers that participated saw a greater appeal to a younger audience as the PizzaBox provided an entertainment value that would be more attractive to a lower age range. The use of physical objects and the idea that they are making a pizza and then seeing the pizza made provides a sense of fantasy that they somehow were involved in the making of the pizza. C5 considered the idea that a younger audience would be more likely to engage with the PizzaBox as they are more exposed to newer technology making them more comfortable around technology in general and would not be afraid to explore the new option of food ordering whereas an older population would be less willing to try something new due to a lack of confidence around technology or a lack of exposure. A conversation with the pizzeria owner also introduced a new concept that we had not thought about when initially coming up with the PizzaBox concept and considering the use cases for the PizzaBox. She pointed out  the idea of educational interactivity. The concept encompasses the idea that our system would introduce children to where meat products such as beef or ham come from. \medskip

P1: \textit{``I was surprised to read that children often don't know that beef comes from a cow or ham comes from a pig, at least with this system they can see the animal and relate that to the food that is being put on their pizza providing a learning experience without them really knowing it.''} \medskip

The conversation with the pizzeria owner continued to discuss how the draw of the PizzaBox being entertaining for younger children will create recurring customers to the pizzeria. The idea that was discussed was as the child finds the PizzaBox fun and engaging they will want to return to use the PizzaBox again, this encourages parents into a regular customer as their child wishes to return the pizza restaurant which has the PizzaBox.

As an opposite to having a greater appeal to a younger audience, we also discussed the PizzaBox being appealing to an older population and we received both negative and positive responses. C1, being one of the younger customers that we interviewed being within the 20 - 30 years bracket, expressed that their older family members are not knowledgeable about current technologies like smartphones/tablets to use food ordering applications or laptops to be able to use a browser to order their food and so an intuitive ordering system like the PizzaBox would appeal to them.\medskip

C1: \textit{``Being a simple system for anyone, including my grandparents that try and avoid all sorts of technology, means they will probably give it shot [if it was presented to them]''}\medskip

While on topic of appealing to an older generation, C3 expressed a different angle for the appeal of the PizzaBox explaining that many of the older populous, including their own family, suffer from medical conditions such as communication problems, hard of hearing or physical disabilities that make it difficult for them to use conventional ordering systems and that the PizzaBox would allow them to easily make a food order without having to be anxious or nervous about approaching a pizzeria waitress. These views reinforce the point that we made in the introduction section that our system would be able to target an audience that suffers from medical conditions such as Parkinson's disease but further research into this would be needed.

Furthermore, C2 and C5 also disagreed that this sort of system would appeal to an older populous.  Their main reasoning for this is that the system had too great a learning curve for someone who is not in touch with technology or that they would need further instruction for first-time use. C5 goes on to say that older people would just prefer to speak to the employee as they will have a general lack of interest in using the system. As well as this C2 expands that without instruction it would lead to confusion and anger giving a negative experience and making them turn to just speaking to the employee or leaving the establishment. Conversation with the Pizzeria owner also expressed a similar concern that if the PizzaBox was too much of a learning curve for any age group it would cause them to ask the employee for further instructions taking time away from the already bust employee which would have negative connotations for the effectiveness of the business.

We also explored whether the PizzaBox would appeal more to an individual or a group of customers. Overall, the customers saw a greater appeal to a group of people as it would create a social experience with each member of the group being able to interact with the PizzaBox\medskip

C2 - \textit{``If it was me and my friends there would be banter around what to order and calling each other out on the food choices that they made...''}\medskip

The continued conversation with C2 is summarised by the PizzaBox creating new social experiences and interaction between friends. In terms of individual interactivity, all customers agreed that the novelties of the PizzaBox would provide a positive effect on their mood upon first or second use of the system but would become stall after multiple uses. 

\subsection{Towards Connected Real-Time Products and Services}
With the recent concerns on data protection and privacy over the past couple of years, we were eager to discuss any concerns customers would have regarding their data and its use within a possible implemented ecosystem surrounding the PizzaBox and a pizzeria. This theme looks at these concerns or lack of. As of the current version of PizzaBox, no data is taken from the customer and the only data that is collected is the data that has been written to the RFID tag and sent to an email address created by us. The questions asked then are based on futuristic versions of the PizzaBox. We also briefly discuss the idea with the pizzeria owner, getting their perspective on how data collected could aid them.

We introduced the idea of data privacy to selected customers to which the responses were very similar, each customer expressed a relaxed view with regards to any data that would be collected from them when using the PizzaBox. C4 expressed that if the PizzaBox was situated at her home they had to add debit or credit card details to the PizzaBox upon initial setup they would not be concerned due to a similar process when ordering food or clothes over a browser.\medskip

C4 - \textit{``It would be the same as adding your card to sites that you buy clothes from; it just makes it a lot easier to order stuff...''}\medskip

When talking to the customers further about possible features the PizzaBox could incorporate, a loyalty scheme was brought up numerous times where they would also place their loyalty token on the PizzaBox to be scan at the same time as the ingredients for some discount or point accumulation. For each customer to be able to join the loyalty scheme they would have to provide some demographic information. This feature was a popular feature and when discussing the issue of providing demographic details, the customers we spoke to maintain a relaxed stance when it came to their data.\medskip

C2 - \textit{``The rewards gained from the loyalty scheme make it worth providing your details to the pizzeria...''}\medskip

Generally, the concern of data collection was a non-issue for the customers that participated in the evaluation. We hypothesise that this could be down to the simplicity of the system and that the customers were unable to see how or what data could be collected. The issue of data collection was brought up while talking to the owner of the pizzeria to which they agreed that the collection of non-identifying data e.g. history of orders from customers using the PizzaBox would be advantageous to them. They expressed how it would aid with stock management as well as identifying possible sales/specials that would be most effective.


To be a successful food ordering system that provides a unique experience, the PizzaBox must provide something that other conventional ordering systems can not provide during each stage of the ordering process, from ingredient selection to committing the order and making the payment. That something extra and unique comes in the form of entertainment and during discussions with participants, the idea that the PizzaBox had an `entertainment value' was very common. As mentioned previously the majority of customers expressed their opinion that this product would be suitable to children for reasons such as creating a fantasy where the child is a pizza maker and they themselves are included in the making of a pizza, a similar experience the child gets when playing with toy oven sets.

Another reason was the introduction to a form of `educational interactivity', where the children can learn of the origins of their food by connecting ingredients to the animal to which they come from just by engaging with the PizzaBox. 

As we discuss later in Section \ref{sec:HealthyEatingandLiving}, the entertainment factor of the PizzaBox also has a great value for the owner of the pizzeria as it encourages the return of customers as they want to use the PizzaBox again to order their food or a child might ask their parents to order from that pizzeria again creating a case were the parents become regular customers and eventually the child will become a regular customer also once they have grown up. Even though they expressed opinions that it would be more suitable for children all the participants agreed that the PizzaBox had a novelty value which would attract all types of people to the product, even if they decided to not engage. Social media  such as Twitter is another dimension that can be incorporated in this design to improve the user experience. For example, Al-Ateeq \cite{AlAteeq2016}  have developed vending machine that uses Twitter API to interact with user's requests.

\subsubsection{Next Generation Customer Relationships}
\label{sec:NextGenerationCustomerRelationships}
As a subsection of the data collection and analytics, it leads us into the use of the PizzaBox for advertisements for the pizzeria or restaurant that it is located in. As previously stated, the data of order history that can be collected is able to aid the pizzeria owner in customising specials, coupons or advertisements for the customers to be able to see on the screen embedded in the PizzaBox. Customers agree that when making their order, being able to see specials on a display would possibly affect their food order depending on the special that was available at the time of food ordering. The pizzeria owner also added that the system would need to be simple to add the specials to the system which would be needed to be considered in revised versions as well as being able to create special coupons for discounts that the customer can use.

\subsection{Healthy Eating and Living}
\label{sec:HealthyEatingandLiving}
It's well known that pizzas are generally an unhealthy food option, so this theme looks at how the PizzaBox is able, in its current revision and in future revisions, to introduce healthier food choices and habits for the customers that use the PizzaBox. Based on the current version we discussed with customers based on their experience that they had with the PizzaBox would they alter their choice of toppings or reduce the number of unhealthy options they chose. C1 expressed that being able to see the ingredients available to them gave them a greater choice and would be more likely to pick a different option than their normal choice of a heavy meat-based pizza.\medskip

C1 - \textit{``Be able to see more options would encourage me to pick something else than the usual meat feast I go for, although might not be a healthier option but I would be more likely to consider it.''}\medskip

Expanding on that view, C4 expressed that actually seeing the ingredients as a 3D form would alter their choice as being able to see the food and the quantity that is being added to the PizzaBox gave the customer a sense that they are adding an unneeded quantity of food and so they would be more willing to take off an ingredient. While talking with the customers only one, C5, disagreed that the PizzaBox would affect their eating habits or their choice of pizza, even after we discussed possible features of future revisions to the PizzaBox. C5 regarded themselves as not a health-conscious person when it came to their food and they are aware that pizzas are generally unhealthy but will accept that as they enjoy a pizza after a shift at work, simply put ``...If I want a pizza, I'll get one''. 

While discussing healthy eating with the customers, we also asked for possible improvements or features that we could add to the PizzaBox to make customers more conscious about their food choices to encourage healthier eating. A popular opinion across most of the customers was a display on the screen of a total calorie count of each ingredient that was on the PizzaBox. Another popular idea that came up was using a traffic light LED system that represented low to high-calorie content as well as being able to warn people easily of allergens in the food that they are ordering. The pizzeria owner explained that allergens can be deadly to some customers so a clear warning must be in place when creating a new ordering system and that an amber warning on the lights and a message on the display would be greatly beneficial to customers that have an allergic reaction to certain foods. Each of the customers agreed that these features would make them more likely to change to healthier options. C2 commented on the LED option as being like the salt/fat/sugar information on packaging in UK Supermarkets.\medskip

C2 - \textit{``...its like when I pick up food in Tesco, if I look on the back of the product and it has a red indicator for fat or salt, I'm instantly more likely to put it down...``}\medskip

C1 had a similar idea that sections of the pizza representing fat, salt etc. would light up to show that there is a high content of that nutritional value in the pizza, essentially providing the customer with a greater break down of nutritional values in the pizza that's created.

Another interesting idea came from C2. They described themselves as being health conscious, (They also explained they were on a cheat day meaning they can eat what they want for the day hence being in pizzeria) and use an app called \textit{`MyFitnessPal'} to input data on food they had consumed that day to help them with gym progression and that integration into these popular health apps would benefit them greatly. 


With each discussion we had with customers that participated we tried to assess how their mood is affected using the PizzaBox. We also discuss with the customer's situations to which a potential customer using the PizzaBox might suffer a negative effect on their mood and why. Throughout the discussions and demonstrations of the PizzaBox, all participants can be seen to be enjoying themselves and using the PizzaBox to create new social cues based on how the PizzaBox works and how the ingredients look. We gained a new participant C4 after they saw C3 laughing during our interactions which shows that the PizzaBox has a draw to be used. As discussed previously, positive effects on the mood of a younger audience would more likely due to the PizzaBox providing a different and fun experience for the child and in turn, the effect on the mood of the parent is improved. This is all based on speculation and discussion with the customers and further research into this would be needed.

While looking at the positive effects we must also consider the negative effects that might possibly arise while using the PizzaBox. We discussed previously that there might not be much of an appeal to an older generation due to a lack of exposure to technology which then leads to confusion and frustration. Again, this is based on speculation and discussion with the participating customers and will need further research. These negative emotions will have considerable effects with the longevity of the use of the PizzaBox so future revisions of the system will need to ensure simplicity and intuitiveness to avoid this.

\subsubsection{Personalisation and Customization}

The discussion of new features for a healthier choice of eating also introduced the idea of being able to add limitations or restrictions to certain aspects of the pizza ordering process e.g. ingredients, quantity or the amount of calories/salt etc. As this does not help to aid a healthier choice but more along the lines of enforcing a healthier choice, we discussed use cases with the customers that could take advantage of this feature. C3 presented the idea that the restriction implementation could be useful to avoid accidental ordering of food that are disliked or are against dietary choices e.g. vegans, vegetarians etc. or that customers avoid due to religious views e.g. Muslim Halal diet. Another suggestion from C3 was that a spouse/parent/friend would be able to set a limitation to how many calories etc. you are able to order per pizza. This would help in making sure children don't order a pizza that a parent deems too unhealthy or it could help with aiding a partner or spouse loose weight by taking away the opportunity to order an unhealthy pizza (or repetitive behaviour). C4 also presented an idea that the restrictions feature would be a popular feature for customers with allergens as they can block any transaction which included an item that would induce an allergic reaction from eating the pizza making ordering a pizza a less stressful and faster experience as they will be able their favourite pizza without worry and having to ask the pizzeria employee (or ring a pizzeria if the PizzaBox is situated in a home environment) to ask for allergen advice. The implementation of such object-based restrictions could follow the ideas proposed by Cervantes-Solis et al. \cite{CervantesSolis2015} where they have demonstrated a use of a goal-directed object manipulation techniques to facilitate the implementation of constraints. The last idea was put forward by C1 who suggested the addition of a biometric scanner for parental authentication when ordering food in a home environment \cite{Kawsar2010}. \medskip

C1 - \textit{``...and my children will be trying to order pizzas whenever they get the chance so adding a fingerprint scanner as I have on my phone, so I have to scan my finger to confirm they can make the order would be hugely beneficial.''}

It is important to note, in this version of the PizzaBox, we were careful not to use the term `smart' PizzaBox. However, as discussed above, there are number of features that we could implement to embed fair bit of `smartness' to this object. In this regard, we see parallels between our work and Farion et al.\cite{Farion2014}, where a smart handbag aims to provide useful reminders (e.g., to take key with you). Reminders to stop repetitive order behaviours or higher amount of calories consumption could be perceived as similar features.

\subsection{Evolution of Food Ordering Systems}
To be able to give us a greater idea of how the use of the PizzaBox in a pizzeria or restaurant environment will be used this theme will look at how the customers and the pizzeria owner compare the PizzaBox to other food ordering systems or processes that are available. Currently, there are four main ordering systems and processes that we compared the PizzaBox with; Web/Online applications, Face to face, making an order over the phone and McDonald's interactive touch screen. When discussing with the customer we randomly selected one or two of these processes to make comparisons. 

Our conversation with the pizzeria owner shone a light on a factor that we already considered multiple times throughout the course of the design process. The observation was the increase in ordering time as the customer would have to physically move the objects over to the base of the PizzaBox as well as the time that is needed for it to scan the ingredients was significantly higher when compared to how quickly a customer can just say their order over the counter. P1 continued to inform that in their experience working in the pizzeria most people will come to the counter with their full order already thought out so it's quicker to talk face to face with the pizzeria employee, a view that C5 also agreed with. P1 also expressed their view that the PizzaBox would not increase the effectiveness of their work for the majority of scenarios but they could see how it would aid them when dealing with groups of size 3+ as they won't need to try and remember orders but can check the email to remind them. 

\begin{figure}[b]
		\vspace{-12pt}
	\centering
	\includegraphics[scale=0.75]{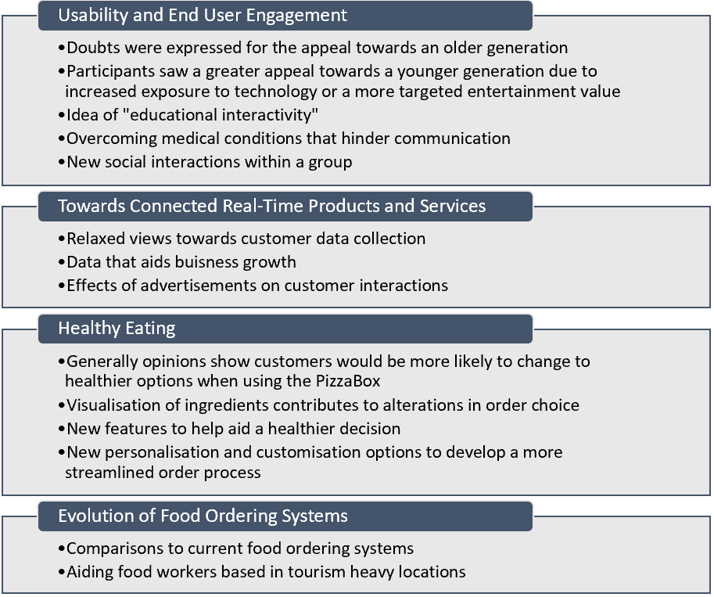}
	\caption{Summary of the Findings}
	\label{fig:discussionsummary}
	
	\vspace{-12pt}
\end{figure}

An interesting comparison around the topic of tourism was made by C3 during our discussions. They explain the idea that the PizzaBox would aid workers in the fast food industry that are situated around a lot of tourism as the physical objects that represent the ingredients and the pizza base are universal which can remove language barriers, something which other ordering processes except the McDonald's touch screen are unable to do. This idea was particularly appealing to the pizzeria owner as the pizzeria at which we were conducting our research was near a motorcycle club that welcomes 100s of motorcyclists from around Europe twice a year which, when they try and order food, can be frustrating for both parties. 

C2 explained during our discussions that they would prefer to avoid social interaction when ordering their food due to low social confidence and that the PizzaBox helps to remove the social interaction that they would usually shy away from. The typically prefer socially interactive objects \cite{Jia2013}. A comparison was made towards the McDonald's touch screen and mobile applications as they both help to remove this hurdle of social anxiety and such C2 compared each of the processes equally. The comparison between the PizzaBox and the McDonald's touchscreen system being equally continued through the majority of our interactions with the customers with C3 even claiming that PizzaBox is a better system as you are able to see what you are ordering in a physical sense instead of just seeing pictures on a screen. C1 was the only customer to make a comparison between the PizzaBox and an online order process with the comparison being in favour of the PizzaBox. C1 explained that often an online process can often be too complicated, or they are easily lost during the process of creating their own pizza but the PizzaBox with its simple process of just adding the ingredients and pressing the order button much more appealing to them. A comparison was made by C2 when discussing the appeal of the PizzaBox to a group of people when attempting to order food over the phone.

C2 - \textit{``When I'm with my friends and we make a food order we always have to write it down or I'm shouting at them to tell me what they want to eat over the phone, and stuff often gets left out. At least with the PizzaBox, it would remove all that fuss...''} \medskip

They make a clear indication that the PizzaBox would be a favourable choice than ordering food over a phone when in a group scenario. The Figure \ref{fig:discussionsummary} summarises our findings.

\section{Conclusion}
The paper introduced an interactive, 3D printed, food ordering system which aims to bring a new experience different from conventional food ordering systems and to provide an entertaining and unique experience when ordering a pizza. PizzaBox was co-designed and went through two prototype testing phases to make sure it was appealing to all age ranges while also being simple and intuitive to use.

The main findings from our research showed how customers would be more willing to make a healthier choice when ordering food using the PizzaBox as it gave an easy visualisation to aspects such as the number of ingredients they wanted to add to the pizza and what they were adding to the pizza. The findings also showed that the PizzaBox can create new social experiences in friendship groups that are collaboratively trying to order a pizza thus having a positive effect on all members within the group. Finally, the PizzaBox provides entertainment value and novelty that has appeal for all age ranges but appeals more to a younger audience by giving them a fantasy experience of making and baking the pizza.

Throughout the interviews with participants, each offered valuable suggestions as to what they see as possible improvements to future revisions of the PizzaBox to help encourage healthy eating behaviours with the most popular suggestion being LEDs that operated a traffic light like system to display a high-calorie count or high salt quantity for example. The 3D printed ingredients were a big hit with each participant, all expressing how easy it was to recognise each ingredient with ease. They also inspired one customer to introduce the idea that the PizzaBox would help employees of fast food establishments that are located in tourist-heavy areas to help with language barriers as the 3D depiction of ingredients are universal for anyone around the world.

Although this was not part of our main assessments, one participant was keen to introduce a use case that was discussed during our introduction to the paper. The use case is that the PizzaBox could aid sufferers of medical conditions such as Parkinson's or conditions which cause communication problems by removing the difficulties they have to overcome when either trying to use a mobile phone or communicate their order face to face with an employee of a pizzeria or other fast food establishments. Due to the time frame in which the testing was conducted this use case remained out of scope but further study into tangible objects to aid medical conditions in the area of food ordering can be considered as desirable.



\appendix


	\section*{Acknowledgement}
%
We acknowledge the funding received by EPSRC PETRAS 2 (EP/S035362/1) and RiR (EP/T517203/1).

\bibliographystyle{ACM-Reference-Format}
\bibliography{library}

%% file: sample-journal.bbl

\begin{thebibliography}{00}


\ifx \showCODEN    \undefined \def \showCODEN     #1{\unskip}     \fi
\ifx \showDOI      \undefined \def \showDOI       #1{#1}\fi
\ifx \showISBNx    \undefined \def \showISBNx     #1{\unskip}     \fi
\ifx \showISBNxiii \undefined \def \showISBNxiii  #1{\unskip}     \fi
\ifx \showISSN     \undefined \def \showISSN      #1{\unskip}     \fi
\ifx \showLCCN     \undefined \def \showLCCN      #1{\unskip}     \fi
\ifx \shownote     \undefined \def \shownote      #1{#1}          \fi
\ifx \showarticletitle \undefined \def \showarticletitle #1{#1}   \fi
\ifx \showURL      \undefined \def \showURL       {\relax}        \fi
\providecommand\bibfield[2]{#2}
\providecommand\bibinfo[2]{#2}
\providecommand\natexlab[1]{#1}
\providecommand\showeprint[2][]{arXiv:#2}

\bibitem[\protect\citeauthoryear{Ackad, Collins, and Kay}{Ackad
  et~al\mbox{.}}{2010}]%
        {Ackad2010}
\bibfield{author}{\bibinfo{person}{Christopher~J. Ackad},
  \bibinfo{person}{Anthony Collins}, {and} \bibinfo{person}{Judy Kay}.}
  \bibinfo{year}{2010}\natexlab{}.
\newblock \showarticletitle{{Switch - Exploring the Design of Application and
  Configuration Switching at Tabletops}}. In \bibinfo{booktitle}{{\em
  Proceedings of the International Conference on Interactive Tabletops and
  Surfaces (ITS'10)}}.
\newblock


\bibitem[\protect\citeauthoryear{Adafruit}{Adafruit}{2019a}]%
        {Adafruit2019}
\bibfield{author}{\bibinfo{person}{Adafruit}.}
  \bibinfo{year}{2019}\natexlab{a}.
\newblock \bibinfo{title}{{Adafruit 9-DOF Accel/Mag/Gyro+Temp Breakout Board -
  LSM9DS1 ID: 3387 - {\$}14.95 : Adafruit Industries, Unique {\&} fun DIY
  electronics and kits}}.
\newblock   (\bibinfo{year}{2019}).
\newblock
\showURL{%
\url{https://www.adafruit.com/product/364https://www.adafruit.com/product/3387}}


\bibitem[\protect\citeauthoryear{Adafruit}{Adafruit}{2019b}]%
        {Arduno}
\bibfield{author}{\bibinfo{person}{Adafruit}.}
  \bibinfo{year}{2019}\natexlab{b}.
\newblock \bibinfo{title}{{ARDUINO UNO WiFi REV2}}.
\newblock   (\bibinfo{year}{2019}).
\newblock
\showURL{%
\url{https://store.arduino.cc/arduino-uno-wifi-rev2https://store.arduino.cc/usa/arduino-uno-wifi-rev2}}


\bibitem[\protect\citeauthoryear{Adafruit}{Adafruit}{2019c}]%
        {Adafruit2019a}
\bibfield{author}{\bibinfo{person}{Adafruit}.}
  \bibinfo{year}{2019}\natexlab{c}.
\newblock \bibinfo{title}{{i2c / SPI character LCD backpack ID: 292 - {\$}9.95
  : Adafruit Industries, Unique {\&} fun DIY electronics and kits}}.
\newblock   (\bibinfo{year}{2019}).
\newblock
\showURL{%
\url{https://www.adafruit.com/product/292}}


\bibitem[\protect\citeauthoryear{Adafruit}{Adafruit}{2019d}]%
        {LCDdisplay}
\bibfield{author}{\bibinfo{person}{Adafruit}.}
  \bibinfo{year}{2019}\natexlab{d}.
\newblock \bibinfo{title}{{Standard LCD 16x2 + extras [white on blue] ID: 181 -
  {\$}9.95 : Adafruit Industries, Unique {\&} fun DIY electronics and kits}}.
\newblock   (\bibinfo{year}{2019}).
\newblock
\showURL{%
\url{https://www.adafruit.com/product/181}}


\bibitem[\protect\citeauthoryear{Al-Ateeq and Al-Khalifa}{Al-Ateeq and
  Al-Khalifa}{2016}]%
        {AlAteeq2016}
\bibfield{author}{\bibinfo{person}{W Al-Ateeq} {and} \bibinfo{person}{H
  Al-Khalifa}.} \bibinfo{year}{2016}\natexlab{}.
\newblock \showarticletitle{{A hands-on approach to the web of things: The
  twitter vending machine experience}}. In \bibinfo{booktitle}{{\em ACM
  International Conference Proceeding Series}}.
\newblock
\showDOI{%
\url{https://doi.org/10.1145/3011141.3011214}}


\bibitem[\protect\citeauthoryear{Apted, Collins, and Kay}{Apted
  et~al\mbox{.}}{2009}]%
        {Apted2009}
\bibfield{author}{\bibinfo{person}{Trent Apted}, \bibinfo{person}{Anthony
  Collins}, {and} \bibinfo{person}{Judy Kay}.} \bibinfo{year}{2009}\natexlab{}.
\newblock \showarticletitle{{Heuristics to Support Design of New Software for
  Interaction at Tabletops}}. In \bibinfo{booktitle}{{\em Proceedings of the
  Workshop on Multitouch and Surface Computing}}.
\newblock


\bibitem[\protect\citeauthoryear{Baraldi, Bimbo, Landucci, Torpei, Cafini,
  Farella, Pieracci, and Benini}{Baraldi et~al\mbox{.}}{2007}]%
        {Baraldi2007}
\bibfield{author}{\bibinfo{person}{Stefano Baraldi},
  \bibinfo{person}{Alberto~Del Bimbo}, \bibinfo{person}{Lea Landucci},
  \bibinfo{person}{Nicola Torpei}, \bibinfo{person}{Omar Cafini},
  \bibinfo{person}{Elisabetta Farella}, \bibinfo{person}{Augusto Pieracci},
  {and} \bibinfo{person}{Luca Benini}.} \bibinfo{year}{2007}\natexlab{}.
\newblock \showarticletitle{{Introducing TANGerINE : A Tangible Interactive
  Natural Environment}}.
\newblock \bibinfo{journal}{{\em Computer\/}} (\bibinfo{year}{2007}).
\newblock
\showISBNx{9781595937018}
\showDOI{%
\url{https://doi.org/10.1145/1291233.1291422}}


\bibitem[\protect\citeauthoryear{{Best Vines}}{{Best Vines}}{2014}]%
        {PizzaHut}
\bibfield{author}{\bibinfo{person}{{Best Vines}}.}
  \bibinfo{year}{2014}\natexlab{}.
\newblock \bibinfo{title}{{How To Properly Order Pizza (Pizza Hut Interactive
  Concept Table) - YouTube}}.
\newblock   (\bibinfo{year}{2014}).
\newblock
\showURL{%
\url{https://www.youtube.com/watch?v=8mFZEKPSyWw}}


\bibitem[\protect\citeauthoryear{Cervantes-Solis, Baber, Khattab, and
  Mitch}{Cervantes-Solis et~al\mbox{.}}{2015}]%
        {CervantesSolis2015}
\bibfield{author}{\bibinfo{person}{J.~Waldo Cervantes-Solis},
  \bibinfo{person}{Chris Baber}, \bibinfo{person}{Ahmad Khattab}, {and}
  \bibinfo{person}{Roman Mitch}.} \bibinfo{year}{2015}\natexlab{}.
\newblock \showarticletitle{{Rule and theme discovery in human interactions
  with an 'internet of things'}}.
\newblock
\showDOI{%
\url{https://doi.org/10.1145/2783446.2783565}}


\bibitem[\protect\citeauthoryear{{Children Interactive game}}{{Children
  Interactive game}}{[n. d.]}]%
        {PizzaPop}
\bibfield{author}{\bibinfo{person}{{Children Interactive game}}.}
  \bibinfo{year}{[n. d.]}\natexlab{}.
\newblock \bibinfo{title}{{Pizza Pop}}.
\newblock   (\bibinfo{year}{[n. d.]}).
\newblock
\showURL{%
\url{http://hougang-street-93.singaporelisted.com/kids-products-toys/children-interactive-game-pizza-pop}}


\bibitem[\protect\citeauthoryear{de~Freitas, Nebeling, Chen, Yang, {Karthikeyan
  Ranithangam}, and Dey}{de~Freitas et~al\mbox{.}}{2016}]%
        {DeFreitas2016}
\bibfield{author}{\bibinfo{person}{Adrian~A de Freitas},
  \bibinfo{person}{Michael Nebeling}, \bibinfo{person}{Xiang~'Anthony' Chen},
  \bibinfo{person}{Junrui Yang}, \bibinfo{person}{Akshaye Shreenithi~Kirupa
  {Karthikeyan Ranithangam}}, {and} \bibinfo{person}{Anind~K Dey}.}
  \bibinfo{year}{2016}\natexlab{}.
\newblock \showarticletitle{{Snap-To-It: A User-Inspired Platform for
  Opportunistic Device Interactions}}. In \bibinfo{booktitle}{{\em Proceedings
  of the 2016 CHI Conference on Human Factors in Computing Systems}}.
\newblock
\showISBNx{978-1-4503-3362-7}
\showDOI{%
\url{https://doi.org/10.1145/2858036.2858177}}


\bibitem[\protect\citeauthoryear{Dixon, Kimes, and Verma}{Dixon
  et~al\mbox{.}}{2009}]%
        {Dixon2009a}
\bibfield{author}{\bibinfo{person}{Michael Dixon}, \bibinfo{person}{Sheryl~E
  Kimes}, {and} \bibinfo{person}{Rohit Verma}.}
  \bibinfo{year}{2009}\natexlab{}.
\newblock \showarticletitle{{Customer Preferences for Restaurant Technology
  Innovations}}.
\newblock \bibinfo{journal}{{\em Cornell Hospitality Report\/}}
  \bibinfo{volume}{9}, \bibinfo{number}{7} (\bibinfo{year}{2009}),
  \bibinfo{pages}{6--16}.
\newblock
\showURL{%
\url{http://scholarship.sha.cornell.edu/chrpubs}}


\bibitem[\protect\citeauthoryear{Domino's}{Domino's}{2018}]%
        {Dominos2018}
\bibfield{author}{\bibinfo{person}{Domino's}.} \bibinfo{year}{2018}\natexlab{}.
\newblock \bibinfo{title}{{Domino's AnyWare}}.
\newblock   (\bibinfo{year}{2018}).
\newblock
\showURL{%
\url{https://anyware.dominos.com/}}


\bibitem[\protect\citeauthoryear{Dubno and Dubno}{Dubno and Dubno}{1982}]%
        {Dubno1982}
\bibfield{author}{\bibinfo{person}{Michael Dubno} {and} \bibinfo{person}{Daniel
  Dubno}.} \bibinfo{year}{1982}\natexlab{}.
\newblock \bibinfo{title}{{Food service ordering terminal with video game
  capability}}.
\newblock   (\bibinfo{date}{dec} \bibinfo{year}{1982}),
  \bibinfo{numpages}{8}~pages.
\newblock
\showURL{%
\url{https://patents.google.com/patent/US4722053A/en}}


\bibitem[\protect\citeauthoryear{Farion and Purver}{Farion and Purver}{2014a}]%
        {Farion}
\bibfield{author}{\bibinfo{person}{Christine Farion} {and}
  \bibinfo{person}{Matthew Purver}.} \bibinfo{year}{2014}\natexlab{a}.
\newblock \showarticletitle{{Did you pack your keys?}}
  \bibinfo{pages}{539--542}.
\newblock
\showISBNx{9781450324748}
\showDOI{%
\url{https://doi.org/10.1145/2559206.2574809}}


\bibitem[\protect\citeauthoryear{Farion and Purver}{Farion and Purver}{2014b}]%
        {Farion2014}
\bibfield{author}{\bibinfo{person}{Christine Farion} {and}
  \bibinfo{person}{Matthew Purver}.} \bibinfo{year}{2014}\natexlab{b}.
\newblock \showarticletitle{{Did You Pack Your Keys?: Smart Objects and
  Forgetfulness}}.
\newblock \bibinfo{journal}{{\em CHI '14 Extended Abstracts on Human Factors in
  Computing Systems\/}} (\bibinfo{year}{2014}).
\newblock
\showISBNx{978-1-4503-2474-8}
\showDOI{%
\url{https://doi.org/10.1145/2559206.2574809}}


\bibitem[\protect\citeauthoryear{Girouard, Solovey, Hirshfield, Ecott, Shaer,
  and Jacob}{Girouard et~al\mbox{.}}{2007}]%
        {Girouard2007}
\bibfield{author}{\bibinfo{person}{Audrey Girouard},
  \bibinfo{person}{Erin~Treacy Solovey}, \bibinfo{person}{Leanne~M.
  Hirshfield}, \bibinfo{person}{Stacey Ecott}, \bibinfo{person}{Orit Shaer},
  {and} \bibinfo{person}{Robert J.~K. Jacob}.} \bibinfo{year}{2007}\natexlab{}.
\newblock \showarticletitle{{Smart Blocks : A Tangible Mathematical
  Manipulative}}.
\newblock \bibinfo{journal}{{\em Proceedings of the 1st international
  conference on Tangible and embedded interaction\/}} (\bibinfo{year}{2007}).
\newblock
\showISBNx{9781595936196}
\showDOI{%
\url{https://doi.org/10.1145/1226969.1227007}}


\bibitem[\protect\citeauthoryear{Jamieson}{Jamieson}{2015}]%
        {Jamieson2015}
\bibfield{author}{\bibinfo{person}{Sophie Jamieson}.}
  \bibinfo{year}{2015}\natexlab{}.
\newblock \bibinfo{title}{{Domino's sales climb as customers order pizza by app
  - Telegraph}}.
\newblock   (\bibinfo{year}{2015}).
\newblock
\showURL{%
\url{https://www.telegraph.co.uk/finance/newsbysector/epic/dom/11767078/Dominos-sales-climb-as-customers-order-pizza-by-app.html}}


\bibitem[\protect\citeauthoryear{Jenkins}{Jenkins}{2015}]%
        {Jenkins2015}
\bibfield{author}{\bibinfo{person}{Tom Jenkins}.}
  \bibinfo{year}{2015}\natexlab{}.
\newblock \showarticletitle{{Designing the "Things" of the IoT}}. In
  \bibinfo{booktitle}{{\em Proceedings of the Ninth International Conference on
  Tangible, Embedded, and Embodied Interaction - TEI '14}}.
\newblock
\showISBNx{9781450333054}
\showDOI{%
\url{https://doi.org/10.1145/2677199.2691608}}


\bibitem[\protect\citeauthoryear{Jia, Wu, Jung, Shapiro, and Sundar}{Jia
  et~al\mbox{.}}{2013}]%
        {Jia2013}
\bibfield{author}{\bibinfo{person}{Haiyan Jia}, \bibinfo{person}{Mu Wu},
  \bibinfo{person}{Eunhwa Jung}, \bibinfo{person}{Alice Shapiro}, {and}
  \bibinfo{person}{S~Shyam Sundar}.} \bibinfo{year}{2013}\natexlab{}.
\newblock \showarticletitle{{When the Tissue Box Says "Bless You": Using Speech
  to Build Socially Interactive Objects}}. In \bibinfo{booktitle}{{\em CHI '13
  Extended Abstracts on Human Factors in Computing Systems}}.
\newblock
\showISBNx{978-1-4503-1952-2}
\showDOI{%
\url{https://doi.org/10.1145/2468356.2468649}}


\bibitem[\protect\citeauthoryear{Kawsar, Rukzio, and Kortuem}{Kawsar
  et~al\mbox{.}}{2010}]%
        {Kawsar2010}
\bibfield{author}{\bibinfo{person}{Fahim Kawsar}, \bibinfo{person}{Enrico
  Rukzio}, {and} \bibinfo{person}{Gerd Kortuem}.}
  \bibinfo{year}{2010}\natexlab{}.
\newblock \showarticletitle{{An explorative comparison of magic lens and
  personal projection for interacting with smart objects}}. In
  \bibinfo{booktitle}{{\em Proceedings of the 12th International Conference on
  Human Computer Interaction with Mobile Devices and Services}}.
  \bibinfo{pages}{157----160}.
\newblock
\showDOI{%
\url{https://doi.org/10.1145/1851600.1851627}}


\bibitem[\protect\citeauthoryear{Khandelwal and Mazalek}{Khandelwal and
  Mazalek}{2007}]%
        {Khandelwal2007}
\bibfield{author}{\bibinfo{person}{Madhur Khandelwal} {and}
  \bibinfo{person}{Ali Mazalek}.} \bibinfo{year}{2007}\natexlab{}.
\newblock \showarticletitle{{Teaching table: a tangible mentor for pre-k math
  education}}.
\newblock \bibinfo{journal}{{\em 1st International Conference on Tangible and
  Embedded Interaction (TEI'07)\/}} (\bibinfo{year}{2007}).
\newblock
\showISBNx{9781595936196}
\showDOI{%
\url{https://doi.org/10.1145/1226969.1227009}}


\bibitem[\protect\citeauthoryear{Kubicki, Wolff, Lepreux, and Kolski}{Kubicki
  et~al\mbox{.}}{2015}]%
        {Kubicki2015}
\bibfield{author}{\bibinfo{person}{S{\'{e}}bastien Kubicki},
  \bibinfo{person}{Marion Wolff}, \bibinfo{person}{Sophie Lepreux}, {and}
  \bibinfo{person}{Christophe Kolski}.} \bibinfo{year}{2015}\natexlab{}.
\newblock \showarticletitle{{RFID interactive tabletop application with
  tangible objects: exploratory study to observe young children' behaviors}}.
\newblock \bibinfo{journal}{{\em Personal and Ubiquitous Computing\/}}
  \bibinfo{volume}{19}, \bibinfo{number}{8} (\bibinfo{date}{dec}
  \bibinfo{year}{2015}), \bibinfo{pages}{1259--1274}.
\newblock
\showISSN{16174909}
\showDOI{%
\url{https://doi.org/10.1007/s00779-015-0891-7}}


\bibitem[\protect\citeauthoryear{Lavenant}{Lavenant}{2018}]%
        {Lavenant2018}
\bibfield{author}{\bibinfo{person}{Cyril Lavenant}.}
  \bibinfo{year}{2018}\natexlab{}.
\newblock \bibinfo{title}{{The NPD Group}}.
\newblock   (\bibinfo{year}{2018}).
\newblock


\bibitem[\protect\citeauthoryear{Mazalek and Winegarden}{Mazalek and
  Winegarden}{2009}]%
        {Mazalek2009}
\bibfield{author}{\bibinfo{person}{Ali Mazalek} {and} \bibinfo{person}{Claudia
  Winegarden}.} \bibinfo{year}{2009}\natexlab{}.
\newblock \showarticletitle{{Architales: physical/digital co-design of an
  interactive story table}}. In \bibinfo{booktitle}{{\em Proceedings of the 3rd
  International Conference on Tangible and Embedded Interaction}}.
\newblock
\showISBNx{9781605584935}


\bibitem[\protect\citeauthoryear{Miles, Huberman, and Saldaña}{Miles
  et~al\mbox{.}}{2013}]%
        {Miles2013}
\bibfield{author}{\bibinfo{person}{Matthew~B. Miles}, \bibinfo{person}{A.~M.
  Huberman}, {and} \bibinfo{person}{Johnny. Saldaña}.}
  \bibinfo{year}{2013}\natexlab{}.
\newblock \bibinfo{booktitle}{{\em {Qualitative data analysis : a methods
  sourcebook}}}.
\newblock 381 pages.
\newblock
\showISBNx{1452257876}


\bibitem[\protect\citeauthoryear{Moraiti, {Vanden Abeele}, Vanroye, and
  Geurts}{Moraiti et~al\mbox{.}}{2015}]%
        {Moraiti2015}
\bibfield{author}{\bibinfo{person}{Argyro Moraiti}, \bibinfo{person}{Vero
  {Vanden Abeele}}, \bibinfo{person}{Erwin Vanroye}, {and} \bibinfo{person}{Luc
  Geurts}.} \bibinfo{year}{2015}\natexlab{}.
\newblock \showarticletitle{{Empowering Occupational Therapists with a
  DIY-toolkit for Smart Soft Objects}}. In \bibinfo{booktitle}{{\em Proceedings
  of the Ninth International Conference on Tangible, Embedded, and Embodied
  Interaction - TEI '14}}. \bibinfo{publisher}{ACM Press},
  \bibinfo{address}{New York, New York, USA}, \bibinfo{pages}{387--394}.
\newblock
\showISBNx{9781450333054}
\showISSN{19302975}
\showDOI{%
\url{https://doi.org/10.1145/2677199.2680598}}


\bibitem[\protect\citeauthoryear{Ng, Kandala, Marie-Foley, Lo, Steenson, and
  Lee}{Ng et~al\mbox{.}}{2016}]%
        {Ng2016}
\bibfield{author}{\bibinfo{person}{Rachel~S Ng}, \bibinfo{person}{Raghavendra
  Kandala}, \bibinfo{person}{Sarah Marie-Foley}, \bibinfo{person}{Dixon Lo},
  \bibinfo{person}{Molly~Wright Steenson}, {and} \bibinfo{person}{Austin~S
  Lee}.} \bibinfo{year}{2016}\natexlab{}.
\newblock \showarticletitle{{Expressing Intent: An Exploration of Rich
  Interactions}}. In \bibinfo{booktitle}{{\em Proceedings of the TEI '16: Tenth
  International Conference on Tangible, Embedded, and Embodied Interaction}}.
\newblock
\showISBNx{978-1-4503-3582-9}
\showDOI{%
\url{https://doi.org/10.1145/2839462.2856526}}


\bibitem[\protect\citeauthoryear{Nicenboim, Giaccardi, and
  Schouwenaar}{Nicenboim et~al\mbox{.}}{2018}]%
        {Nicenboim2018}
\bibfield{author}{\bibinfo{person}{Iohanna Nicenboim}, \bibinfo{person}{Elisa
  Giaccardi}, {and} \bibinfo{person}{Marcel Schouwenaar}.}
  \bibinfo{year}{2018}\natexlab{}.
\newblock \showarticletitle{{Everyday Entanglements Of The Connected Home}}.
\newblock
\showDOI{%
\url{https://doi.org/10.1145/3170427.3186596}}


\bibitem[\protect\citeauthoryear{Okada, Ueki, Jonasson, Yamanouchi, Norlin,
  Sunahara, Formo, Anneroth, and Inakage}{Okada et~al\mbox{.}}{2016}]%
        {Okada2016}
\bibfield{author}{\bibinfo{person}{Miyo Okada}, \bibinfo{person}{Atsuro Ueki},
  \bibinfo{person}{Niclas Jonasson}, \bibinfo{person}{Masato Yamanouchi},
  \bibinfo{person}{Cristian Norlin}, \bibinfo{person}{Hideki Sunahara},
  \bibinfo{person}{Joakim Formo}, \bibinfo{person}{Mikael Anneroth}, {and}
  \bibinfo{person}{Masa Inakage}.} \bibinfo{year}{2016}\natexlab{}.
\newblock \showarticletitle{{Autonomous Cooperation of Social Things: Designing
  a System for Things with Unique Personalities in IoT}}. In
  \bibinfo{booktitle}{{\em Proceedings of the 6th International Conference on
  the Internet of Things}}.
\newblock
\showISBNx{978-1-4503-4814-0}
\showDOI{%
\url{https://doi.org/10.1145/2991561.2991574}}


\bibitem[\protect\citeauthoryear{Patten, Patten, Ishii, Ishii, Hines, Hines,
  Pangaro, and Pangaro}{Patten et~al\mbox{.}}{2001}]%
        {Patten}
\bibfield{author}{\bibinfo{person}{James Patten}, \bibinfo{person}{James
  Patten}, \bibinfo{person}{Hiroshi Ishii}, \bibinfo{person}{Hiroshi Ishii},
  \bibinfo{person}{Jim Hines}, \bibinfo{person}{Jim Hines},
  \bibinfo{person}{Gian Pangaro}, {and} \bibinfo{person}{Gian Pangaro}.}
  \bibinfo{year}{2001}\natexlab{}.
\newblock \showarticletitle{{Sensetable: A Wireless Object Tracking Platform
  for Tangible User Interfaces}}.
\newblock \bibinfo{journal}{{\em Proceedings of the SIGCHI conference on Human
  factors in computing systems\/}} (\bibinfo{year}{2001}),
  \bibinfo{pages}{253--260}.
\newblock
\showISBNx{1581133278}
\showDOI{%
\url{https://doi.org/10.1145/365024.365112}}


\bibitem[\protect\citeauthoryear{Perera, Liu, and Jayawardena}{Perera
  et~al\mbox{.}}{2015}]%
        {Perera2015a}
\bibfield{author}{\bibinfo{person}{Charith Perera}, \bibinfo{person}{Chi~Harold
  Liu}, {and} \bibinfo{person}{Srimal Jayawardena}.}
  \bibinfo{year}{2015}\natexlab{}.
\newblock \showarticletitle{{The Emerging Internet of Things Marketplace from
  an Industrial Perspective: A Survey}}.
\newblock \bibinfo{journal}{{\em IEEE Transactions on Emerging Topics in
  Computing\/}} \bibinfo{volume}{3}, \bibinfo{number}{4}
  (\bibinfo{year}{2015}), \bibinfo{pages}{585--598}.
\newblock
\showISBNx{2168-6750 VO - 3}
\showISSN{21686750}


\bibitem[\protect\citeauthoryear{Perera, Zaslavsky, Christen, and
  Georgakopoulos}{Perera et~al\mbox{.}}{2014}]%
        {ZMP007}
\bibfield{author}{\bibinfo{person}{Charith Perera}, \bibinfo{person}{Arkady
  Zaslavsky}, \bibinfo{person}{Peter Christen}, {and}
  \bibinfo{person}{Dimitrios Georgakopoulos}.} \bibinfo{year}{2014}\natexlab{}.
\newblock \showarticletitle{{Context Aware Computing for The Internet of
  Things: A Survey}}.
\newblock \bibinfo{journal}{{\em Communications Surveys Tutorials, IEEE\/}}
  \bibinfo{volume}{16}, \bibinfo{number}{1} (\bibinfo{year}{2014}),
  \bibinfo{pages}{414 -- 454}.
\newblock
\showISSN{1553-877X}


\bibitem[\protect\citeauthoryear{Richards}{Richards}{2014}]%
        {Richards}
\bibfield{author}{\bibinfo{person}{Lyn Richards}.}
  \bibinfo{year}{2014}\natexlab{}.
\newblock \bibinfo{booktitle}{{\em {Handling qualitative data : a practical
  guide}}}.
\newblock 236 pages.
\newblock
\showISBNx{9781446276068}


\bibitem[\protect\citeauthoryear{Riddiford, Potter, and Hunwick}{Riddiford
  et~al\mbox{.}}{2007}]%
        {Riddiford2007}
\bibfield{author}{\bibinfo{person}{Martin Riddiford}, \bibinfo{person}{Daniel
  Potter}, {and} \bibinfo{person}{Noel Hunwick}.}
  \bibinfo{year}{2007}\natexlab{}.
\newblock \bibinfo{title}{{Interactive Food and Drink Ordering System}}.
\newblock   (\bibinfo{date}{dec} \bibinfo{year}{2007}),
  \bibinfo{numpages}{28}~pages.
\newblock
\showURL{%
\url{https://patents.google.com/patent/US20100106607A1/en?oq=US+2010}}


\bibitem[\protect\citeauthoryear{Sciuto, Saini, Forlizzi, and Hong}{Sciuto
  et~al\mbox{.}}{2018}]%
        {Sciuto2018}
\bibfield{author}{\bibinfo{person}{Alex Sciuto}, \bibinfo{person}{Arnita
  Saini}, \bibinfo{person}{Jodi Forlizzi}, {and} \bibinfo{person}{Jason~I.
  Hong}.} \bibinfo{year}{2018}\natexlab{}.
\newblock \showarticletitle{{Hey Alexa, What's Up? A Mixed-Methods Studies of
  In-Home Conversational Agent Usage}}. In \bibinfo{booktitle}{{\em Proceedings
  of the 2018 on Designing Interactive Systems Conference 2018 - DIS '18}}.
\newblock
\showISBNx{9781450351980}
\showDOI{%
\url{https://doi.org/10.1145/3196709.3196772}}


\bibitem[\protect\citeauthoryear{Sheffield}{Sheffield}{2015}]%
        {Sheffield2015}
\bibfield{author}{\bibinfo{person}{Hazel Sheffield}.}
  \bibinfo{year}{2015}\natexlab{}.
\newblock \bibinfo{title}{{Domino's Pizza sales up 19{\%} thanks to mobile app
  | The Independent}}.
\newblock   (\bibinfo{year}{2015}).
\newblock
\showURL{%
\url{https://www.independent.co.uk/news/business/news/domino-s-pizza-sales-up-19-thanks-to-mobile-app-a6693761.html}}


\bibitem[\protect\citeauthoryear{Soro, Brereton, Dema, Oliver, Chai, and
  Ambe}{Soro et~al\mbox{.}}{2018}]%
        {Soro2018}
\bibfield{author}{\bibinfo{person}{Alessandro Soro}, \bibinfo{person}{Margot
  Brereton}, \bibinfo{person}{Tshering Dema}, \bibinfo{person}{Jessica~L
  Oliver}, \bibinfo{person}{Min~Zhen Chai}, {and} \bibinfo{person}{Aloha
  May~Hufana Ambe}.} \bibinfo{year}{2018}\natexlab{}.
\newblock \showarticletitle{{The Ambient Birdhouse}}. In
  \bibinfo{booktitle}{{\em Proceedings of the 2018 CHI Conference on Human
  Factors in Computing Systems - CHI '18}}. \bibinfo{publisher}{ACM Press},
  \bibinfo{address}{New York, New York, USA}, \bibinfo{pages}{1--13}.
\newblock
\showISBNx{9781450356206}
\showDOI{%
\url{https://doi.org/10.1145/3173574.3173971}}


\bibitem[\protect\citeauthoryear{SparkFun}{SparkFun}{2019}]%
        {SparkSRTR}
\bibfield{author}{\bibinfo{person}{SparkFun}.} \bibinfo{year}{2019}\natexlab{}.
\newblock \bibinfo{title}{{SparkFun Simultaneous RFID Reader - M6E Nano -
  SEN-14066 - SparkFun Electronics}}.
\newblock   (\bibinfo{year}{2019}).
\newblock
\showURL{%
\url{https://www.sparkfun.com/products/14066}}


\bibitem[\protect\citeauthoryear{Storz, Kanellopoulos, Fraas, and Eibl}{Storz
  et~al\mbox{.}}{2014}]%
        {Storz2014}
\bibfield{author}{\bibinfo{person}{Michael Storz}, \bibinfo{person}{Kalja
  Kanellopoulos}, \bibinfo{person}{Claudia Fraas}, {and}
  \bibinfo{person}{Maximilian Eibl}.} \bibinfo{year}{2014}\natexlab{}.
\newblock \showarticletitle{{ComforTable: A Tabletop for Relaxed and Playful
  Interactions in Museums}}. In \bibinfo{booktitle}{{\em Proceedings of the
  Ninth ACM International Conference on Interactive Tabletops and Surfaces}}
  {\em (\bibinfo{series}{ITS '14})}. \bibinfo{publisher}{ACM},
  \bibinfo{address}{New York, NY, USA}, \bibinfo{pages}{447--450}.
\newblock
\showISBNx{978-1-4503-2587-5}
\showDOI{%
\url{https://doi.org/10.1145/2669485.2669531}}


\bibitem[\protect\citeauthoryear{STROM}{STROM}{2017}]%
        {STROM2017}
\bibfield{author}{\bibinfo{person}{STEPHANIE STROM}.}
  \bibinfo{year}{2017}\natexlab{}.
\newblock \bibinfo{title}{{McDonald's Introduces Screen Ordering and Table
  Service}}.
\newblock   (\bibinfo{year}{2017}).
\newblock
\showURL{%
\url{https://www.nytimes.com/2016/11/18/business/mcdonalds-introduces-screen-ordering-and-table-service.html}}


\bibitem[\protect\citeauthoryear{Tan, Lee, Khor, Goh, Tan, and Lew}{Tan
  et~al\mbox{.}}{2010}]%
        {Tan2010}
\bibfield{author}{\bibinfo{person}{YongChai Tan}, \bibinfo{person}{KienLoong
  Lee}, \bibinfo{person}{ZhiChao Khor}, \bibinfo{person}{KaeVin Goh},
  \bibinfo{person}{KhimLeng Tan}, {and} \bibinfo{person}{BentFei Lew}.}
  \bibinfo{year}{2010}\natexlab{}.
\newblock \showarticletitle{{Automated food ordering system with interactive
  user interface approach}}. In \bibinfo{booktitle}{{\em 2010 IEEE Conference
  on Robotics, Automation and Mechatronics, RAM 2010}}.
\newblock
\showISBNx{9781424465033}
\showDOI{%
\url{https://doi.org/10.1109/RAMECH.2010.5513147}}


\bibitem[\protect\citeauthoryear{Weiss, Schwarz, Jakubowski, and
  Borchers}{Weiss et~al\mbox{.}}{2010}]%
        {Weiss2010}
\bibfield{author}{\bibinfo{person}{Malte Weiss}, \bibinfo{person}{Florian
  Schwarz}, \bibinfo{person}{Simon Jakubowski}, {and} \bibinfo{person}{Jan
  Borchers}.} \bibinfo{year}{2010}\natexlab{}.
\newblock \showarticletitle{{Madgets: Actuating Widgets on Interactive
  Tabletops}}. In \bibinfo{booktitle}{{\em Proceedings of the 23nd annual ACM
  symposium on User interface software and technology - UIST '10}}.
\newblock
\showISBNx{9781450302715}
\showDOI{%
\url{https://doi.org/10.1145/1866029.1866075}}


\bibitem[\protect\citeauthoryear{Zolfasharifard}{Zolfasharifard}{2014}]%
        {Zolfasharifard2014}
\bibfield{author}{\bibinfo{person}{Ellie Zolfasharifard}.}
  \bibinfo{year}{2014}\natexlab{}.
\newblock \bibinfo{title}{{Pizza Hut reveals interactive table concept that
  lets you design your perfect pie | Daily Mail Online}}.
\newblock   (\bibinfo{year}{2014}).
\newblock
\showURL{%
\url{https://www.dailymail.co.uk/sciencetech/article-2573164/Pizza-Hut-reveals-interactive-table-concept-lets-design-perfect-pie.html}}


\end{thebibliography}
